\documentclass[preprint,3p,authoryear,a4paper]{elsarticle}
\usepackage{graphicx}
\usepackage{parskip}
\usepackage{xcolor}
\usepackage{dsfont}
\usepackage{amsmath}
\usepackage{amssymb}
\usepackage{amsfonts}
\usepackage{amsbsy}
\usepackage{amsthm}
\usepackage{varioref}
\usepackage{array}
\usepackage{booktabs} % TBV
\usepackage{verbatim} % TBV
\usepackage[english]{babel}
\usepackage{float}
\usepackage{subfigure}
\usepackage{epsfig}
\usepackage{multirow}
\usepackage{pdflscape}
\usepackage{mathrsfs}
\usepackage{lscape}
\usepackage{ulem}
\usepackage{enumerate}
\usepackage{enumitem}
\usepackage{caption}
\usepackage{url}

\newcolumntype{C}[1]{>{\centering\let\newline\\\arraybackslash\hspace{0pt}}m{#1}}
\renewcommand\appendix{\par
\setcounter{section}{0}%
\setcounter{subsection}{0}%
\setcounter{table}{0}
\setcounter{table}{0}
\setcounter{figure}{0}
\gdef\thetable{\Alph{table}}
\gdef\thefigure{\Alph{figure}}
\gdef\thesection{\Alph{section}}
\setcounter{section}{0}}

\DeclareMathOperator{\Var}{Var}

\newcounter{arclist}

\newcounter{arcenum}

\newcommand{\MV}[3]{\operatorname{MV}^{(\text{#1})}\left(#2;#3\right)}

\newcommand{\Ben}[2]{\mathfrak{B}^{(\text{#1})}\left(#2\right)}

\newcommand{\prem}[3]{\mathcal{P}^{(\text{#1})}\left(#2;#3\right)}

\begin{document}

\begin{frontmatter}

\title{
When Indemnity Insurance Fails: \\ Parametric Coverage under Binding Budget and Risk Constraints
}

\author[UM]{Benjamin Avanzi\corref{cor}}
\ead{b.avanzi@unimelb.edu.au}

\author[UoL]{Debbie Kusch Falden}
\ead{dkfalden@liverpool.ac.uk}

\author[Cop]{Mogens Steffensen}
\ead{mogens@math.ku.dk}

\cortext[cor]{Corresponding author. }

\address[UM]{Department of Economics, Faculty of Business and Economics \\ University of Melbourne, Melbourne VIC 3010, Australia}

\address[UoL]{Department of Mathematical Sciences, University of Liverpool\\ L69 7ZL, Liverpool, UK }

\address[Cop]{Department of Mathematical Sciences, University of Copenhagen \\ DK-2100 Copenhagen, Denmark}

\begin{abstract}

%%%
%standard abstract
In high-risk environments, traditional indemnity insurance is often unaffordable or ineffective, despite its well-known optimality under expected utility. We compare excess-of-loss indemnity insurance with parametric insurance within a common mean-variance framework, allowing for fixed costs, heterogeneous premium loadings, and binding budget constraints. Motivated by the disaster insurance and risk-sharing literature, we show that, once these realistic frictions are introduced,  parametric insurance can yield higher welfare for risk-averse individuals, even under the same utility objective and without relying on behavioral assumptions. The welfare advantage arises precisely when indemnity insurance becomes impractical (particularly when households face binding premium budgets), and disappears once both contracts are unconstrained. Our results help reconcile classical insurance theory with the growing use of parametric risk transfer in high-risk settings, and rationalize the interest in hybrid designs that combine both indemnity and parametric elements.

\end{abstract}

\begin{keyword}
Insurance design \sep
Disaster risk \sep
Parametric insurance \sep
Affordability constraints \sep
Risk sharing

JEL codes: %http://www.aeaweb.org/journal/jel_class_system.php
C44 \sep %Statistical Decision Theory; Operations Research
C61 \sep %Optimization Techniques; Programming Models; Dynamic Analysis
G22 \sep % G2	Financial Institutions and Services G22	Insurance • Insurance Companies • Actuarial Studies
G32 \sep %Financing Policy; Financial Risk and Risk Management; Capital and Ownership Structure

MSC classes: 
60G51 \sep % Processes with independent increments
93E20 \sep % Optimal stochastic control
91G70 \sep 	%Statistical methods; risk measures [See also 62P05, 62P20] in Actuarial science and mathematical finance
%91G60 \sep 	%Numerical methods (including Monte Carlo methods) in Actuarial science and mathematical finance
62P05 %\sep 	%Applications of statistics to actuarial sciences and financial mathematics
%62H12 %\sep 	%Estimation in multivariate analysis
91B30 %\sep % Risk theory, insurance

%Subject Categories: %http://www.elsevier.com/authored_subject_sections/S04/misc/subject_cat.htm
%IM13 \sep %ruin and other stability criteria
%IE50 %\sep %finance, general and miscellaneous
%IE53 \sep investment

\end{keyword}

\end{frontmatter}

\newtheorem{remark}{Remark}[section]
\numberwithin{equation}{section}

\section{Introduction}

Insurance markets are increasingly strained in regions exposed to low-frequency, high-severity risks such as floods, wildfires, and cyclones. In many such areas, indemnity insurance has become prohibitively expensive or altogether unavailable, as rising hazard intensity, capital requirements, and fixed costs push premiums and deductibles to levels that leave households effectively uninsured, \citep[e.g.,][]{Ko19,NeKoNeDaKu21, SwissRe25}. A key practical limitation is that household indemnity insurance is typically written on a full-value (sum-insured) basis: the premium is charged to cover the entire exposure, and the main lever to restore affordability is to raise the deductible. In high-risk areas, this mechanism can push deductibles to economically (and practically) unrealistic levels (e.g., tens or hundreds of thousands of dollars), rendering the contract theoretically available but impractical, and delivering negligible effective protection. By contrast, parametric insurance is inherently scalable: households can purchase a smaller, explicitly bounded layer of protection (e.g., a fixed payment per event) without having to insure the full loss. Crucially, parametric coverage scales by adding first-dollar protection, whereas scaling indemnity insurance requires removing first-dollar coverage through higher deductibles, quickly rendering partial indemnity ineffective. Recent empirical evidence reinforces the centrality of these design choices for
households: \citet{JuJu26}, analyzing millions of US homeowner's insurance
contracts, document that deductibles rather than coverage limits are the
economically binding margin shaping household risk exposure. Throughout, our focus is on \textit{household property insurance}, in which the loss exposure is bounded above by the sum insured $L$ on the property.

Recent developments in flood and wildfire insurance markets illustrate this (un)insurability trend: insurers have withdrawn from parts of the California residential market, flood insurance take-up remains persistently low in high-risk zones, and coverage gaps are expanding even in advanced economies \citep[e.g.,][]{KouskyCooke2012,JaBeCh23,undrr-gar-2025}. Yet when disasters occur, losses are rarely borne privately alone. Instead, they are often socialized ex post through government relief, reconstruction grants, and implicit guarantees, a pattern that has been widely documented in the disaster insurance literature \citep{KouskyKunreuther2014,Kousky2018}. This growing disconnect between private insurance provision and public risk-bearing raises a fundamental economic question: when full indemnification is no longer feasible, what forms of risk transfer can still improve individual welfare?

Classical insurance theory provides a clear benchmark. Under broad and well-understood conditions, an excess-of-loss indemnity contract maximizes the expected utility of a risk-averse agent when premiums are loaded proportionally to expected losses \citep{Arrow1971,Raviv1979}. This result has shaped both theoretical and practical views of optimal insurance design for decades. However, it relies on assumptions that are increasingly violated in high-risk environments, as seen in \cite{KouskyCooke2012, Fr01}. In particular, it abstracts from differences in risk borne and fixed costs borne by insurers across contract types, as well as binding budget constraints faced by households. When insurance is expensive, the deductible that is still affordable becomes so large that indemnity insurance offers little or no effective protection, even though it remains optimal in a purely formal sense. In such cases, the classical optimality result continues to hold mathematically, but loses much of its economic content.

This paper revisits the optimality of indemnity insurance from an implementation and affordability perspective, by comparing it with parametric insurance. Parametric contracts replace loss-based indemnification with a fixed payment triggered by an objective index, such as the occurrence of a disaster event. While this design introduces basis risk (arising from imperfect loss matching), it eliminates most of the loss adjustment costs, allows for rapid payouts, and materially reduces the risk borne by insurers \citep{ Ko19, FiMaStYo18, KuMeMiSh13, LiKw20}. As a result, parametric insurance can be priced with lower loadings and smaller fixed costs, and can remain feasible in environments where indemnity insurance breaks down. Despite its growing use in sovereign and corporate risk transfer, parametric insurance for households remains limited in scale and adoption, particularly relative to the extensive theoretical and policy literature on index and disaster insurance \citep{Clarke2016,JeBaMu16, CaJaSaSa17}. That said, policy and industry discussions increasingly contemplate parametric household covers for catastrophic perils. 

A growing literature has analyzed index and parametric insurance, particularly in the context of agricultural risk, sovereign disaster finance, and developing economies \citep[e.g.,][]{Clarke2016,ClarkeDercon2016,HuangShi2024,KouskyCooke2012}. The present paper emphasizes basis risk, demand frictions, and the role of parametric instruments in providing partial protection and liquidity when indemnity insurance is unavailable or inefficient. Our paper is complementary to this literature, but distinct in focus and method. Rather than studying index quality, take-up, or behavioral responses, we ask a specific and unresolved theoretical question: when standard indemnity insurance remains formally optimal under classical criteria but becomes economically irrelevant due to fixed costs and budget constraints, can parametric insurance dominate under the same utility objective?

Nevertheless, indemnity insurance is still widely considered the benchmark for effective protection. We show that this benchmark can lose practical relevance once affordability and implementation frictions are taken into account. Using a tractable model of frequency and severity, we compare indemnity and parametric insurance under the same expected mean--variance utility criterion. Losses arrive according to a Poisson process and have censored exponential severity. This specification can capture the salient features of natural disaster risk (low-frequency, high-severity) while allowing for closed-form solutions. The indemnity contract is of excess-of-loss type with a deductible, while the parametric contract pays a fixed amount per event and shares the same trigger, so there is no basis risk on frequency. Importantly, we allow for distinct premium loadings and fixed costs across contract types, reflecting the higher capital requirements, claims management costs, and reconstruction delays associated with indemnity insurance. At the sovereign level, parametric instruments are widely used as part of fiscal layering strategies, where rapid, rules-based payouts provide immediate liquidity after disasters while larger losses are addressed through slower mechanisms. A similar logic applies at the household level: even limited parametric payments can play a critical role in enabling early recovery and reducing reliance on ad hoc public assistance \citep[e.g.,][]{JaMeMa24}.

Our main result is that, once these realistic frictions are introduced, parametric insurance can dominate indemnity insurance under the same utility objective. This dominance arises most clearly when households face binding budget constraints. In such cases, the deductible required to make indemnity insurance affordable may be so large that the contract is economically irrelevant. In contrast, a parametric contract can still provide a modest but meaningful transfer. Even when this transfer covers only a small fraction of the realized loss, it can materially improve welfare by reducing liquidity shortfalls and supporting early recovery. We characterize indifference thresholds between the two designs and show that the welfare advantage of parametric insurance follows a non-monotonic pattern in the available premium budget: it emerges when budgets are small, disappears as indemnity insurance becomes effective, and vanishes entirely once both contracts are unconstrained.

These findings contribute to several strands of the literature. 

Firstly, they complement the classical optimal insurance results by identifying conditions under which their practical relevance breaks down, without abandoning expected-utility-based evaluation. Our analysis complements the classical insurance literature following \citet{Arr63} and \citet{Raviv1979}, which establishes the optimality of deductible insurance under proportional pricing and frictionless markets. Subsequent work shows that the structure of optimal indemnity insurance is preserved or modified under more general environments, including non-linear premium principles and broader contract constraints \citep{HuMaSm83,Bl85,
GoSh96}, and limited-liability settings in which capped deductibles emerge endogenously as optimal designs \citep{BeKoMu25}. Rather than challenging these results, we show how their economic relevance may erode when heterogeneous fixed costs and loadings, as well as budget constraints, are introduced. In this sense, the paper bridges the gap between foundational insurance theory and the observed expansion of parametric risk transfer in high-risk environments. 

Secondly, they add to the economic analysis of parametric insurance. Existing work has examined the demand for index insurance and its welfare implications, particularly in agricultural and low-income settings  \citep[e.g.,][]{Clarke2016, JeBaMu16, CaJaSaSa17, ChMuBaTu17}, as well as the role of basis risk and index quality in contract performance and how improved trigger design can mitigate discrepancies between actual losses and index-based payouts \citep{CuLaPh04, TeWo19, ZhTaWe19, ClBoBrAe18}. By contrast, our analysis abstracts from index optimization and treats the trigger structure as given, focusing instead on how affordability constraints and implementation frictions shape the relative performance of parametric and indemnity insurance. A related strand emphasizes the use of parametric insurance for extreme risks and catastrophe coverage, highlighting challenges of solvency, tail risk, and capital constraints that limit the feasibility of full indemnification in high-risk environments \citep{ClBoBrAe18, LoTh23}. The literature has largely focused on sovereign and corporate applications \citep{Clarke2016,HuSh24}.

Thirdly, they speak directly to current policy debates on disaster risk financing, insurance affordability, and the appropriate role of governments in supporting risk transfer in high-risk areas. Our analysis suggests that parametric insurance should not be viewed merely as an inferior substitute for indemnity insurance, but as a potentially welfare-improving instrument precisely where traditional insurance fails. While recent work has compared parametric and indemnity insurance in environments with informational frictions, solvency constraints, or general equilibrium interactions \citep{FeJoNo22,We24,LoNk25}, and concurrent work \citep{CaSa26} formalises the optimal layering of budget-constrained
sovereigns across reserve funds, contingent credit, and parametric insurance, we show that parametric insurance can outperform indemnity insurance even in the absence of informational imperfections or equilibrium feedbacks, and at the household rather than sovereign level.

The key message is that, once affordability and implementation frictions are taken seriously, parametric insurance can improve welfare precisely in the environments where indemnity insurance becomes economically irrelevant.

To our knowledge, existing economic analyses of parametric insurance do not provide a closed-form welfare comparison with optimal excess-of-loss indemnity insurance under a common objective, nor do they isolate the role of fixed costs and budget constraints in rendering deductible insurance economically ineffective despite its classical optimality. By keeping preferences, pricing principles, and triggers aligned across contract types, our framework isolates a simple but underappreciated mechanism: parametric insurance remains scalable and welfare-improving precisely because it adds first-dollar protection, whereas partial indemnity insurance removes it.

The remainder of the paper is organized as follows. Section~2 introduces the model and utility framework. Section~3 derives the optimal indemnity and parametric contracts in the absence of budget constraints. Section~4 analyzes the comparison under binding premium budgets and presents numerical illustrations. Section~5 discusses policy implications for disaster risk reduction, insurance market design, and public intervention. Section~6 concludes.

\section{Theoretical framework for traditional indemnity and parametric insurance}
\label{sec:framework}

\subsection{Losses and benefits}

We consider an agent who faces aggregate losses $S$ over a year that may stem from multiple events:
\begin{align}
  S=\sum_{i=1}^{N} Y_i,
\end{align}
where $N$ denotes the number of loss-causing events in the year and $Y_i$ are i.i.d.\ severities independent of N.\footnote{Throughout, we speak of an ``event count'' $N$, but the same notation also covers a ``claim count'' interpretation. The distinction matters only when a single event can generate multiple claims; see Remark~\ref{R:eventvsclaim}.}
Let $\Ben{.}{N,Y_1,\dots,Y_N,.}$ be the insurance benefit (indemnity) paid by the insurer against the aggregate loss $S$.

We compare two stylized benefit structures. In the \textit{traditional} indemnity design,
$\Ben{d}{N,Y_1,\dots,Y_N,d}$, the insurer indemnifies the realized loss above a deductible $d$:
\begin{align}
  \Ben{d}{N,Y_1,\dots,Y_N,d}
  = \sum_{i=1}^N (Y_i-d)_+,
  \qquad (a-b)_+ =\max(a-b,0).
\end{align}
In the \textit{parametric} design,
$\Ben{p}{N,Y_1,\dots,Y_N,k}$, the insurer pays a constant amount $k$ per event regardless of the realized severity:
\begin{align}
  \Ben{p}{N,Y_1,\dots,Y_N,k}
  = \sum_{i=1}^N k
  = kN.
\end{align}
Because the parametric payment depends on $N$ but not on the individual severities $Y_i$, we henceforth write $\Ben{p}{N,k}$ to emphasize that its residual uncertainty is driven entirely by the event count. In practice, what constitutes an increment of $N$ is defined by an objective trigger based on event characteristics (e.g.\ flood depth or wind speed), rather than on the monetary loss $Y_i$ itself (e.g.\ the cost of repairing a house).

The parametric structure also has operational implications. Because $\Ben{p}{N,k}$ does not require an assessment of $Y_i$, claims handling can be fast and inexpensive provided the trigger defining $N$ is transparent and verifiable. Two examples illustrate this mechanism. FloodFlash is a parametric flood cover available in the UK and the US: a device attached to the insured property measures flood depth and pays predetermined amounts as a function of observed water height.\footnote{FloodFlash offers products for both individuals and businesses, though marketing materials appear primarily targeted at commercial customers.} Redicova is a parametric cyclone cover available in Australia (mainly in Queensland), where the trigger is defined using wind speed (as measured by the Australian Bureau of Meteorology) and location; typical insureds include banana farmers. These operational differences motivate allowing the premium structures of the two products to differ; see Section~\ref{Pd}--\ref{Pp} below.

A salient difference is that the parametric design can overpay relative to realized damage: if $k>Y_i$, the insured makes a profit on that event. This issue is mitigated in the catastrophic-risk applications that motivate this paper. First, for disasters such as floods, cyclones/hurricanes/typhoons, and bushfires/wildfires, the insured has no meaningful control over the incidence of events, so classical moral-hazard concerns with respect to $N$ are limited.\footnote{By contrast, moral hazard can be more salient under indemnity: generous coverage may weaken incentives to mitigate severity (e.g.\ by undertaking protective measures), and slow or contested loss adjustment can create opportunities for exaggeration or fraud in large-loss settings.} Second, the demand for parametric cover is most relevant precisely when full traditional insurance is unaffordable or unavailable because losses are deemed effectively uninsurable; in that regime, severities $Y_i$ are typically so large that they will exceed any affordable per-event payment $k$, making both the likelihood and the magnitude of ``profit'' cases small.

Finally, while we adopt the simplest parametric structure $\Ben{p}{N,k}$, the framework naturally extends to a wider family of \textit{parametric indemnity functions. Commercial parametric contracts cover a broad spectrum of designs: constant-per-event payments (as in the simplest Redicova cyclone covers), piecewise-constant schedules indexed on hazard intensity (as in FloodFlash, where payouts step up with observed flood depth), and continuously variable indemnifications indexed on auxiliary observables (as in agricultural index insurance, where payouts scale with remotely sensed yield loss, rainfall deficit, or vegetation indices; see e.g.\ \citealt{Clarke2016, CaJaSaSa17,MaSc26}). Such generalizations (i) introduce dependence between the trigger and payment levels, complicating moment computations; (ii) reduce basis risk (see Remark~\ref{R:basisrisk}); and (iii) increase expected utility by reducing the volatility of net losses.} Since (ii)--(iii) make parametric insurance \textit{more} attractive, we deliberately focus on the most conservative benchmark $\Ben{p}{N,k}$: any welfare gains established in this setting should be interpreted as lower bounds for designs with richer payout functions.

\begin{remark}[Event-level versus claim-level deductibles]\label{R:eventvsclaim}
Classical results due to \citet{Borch1960,Borch1962,Borch1974} and \citet{Arr63,Arrow1971} show that, under concave preferences and linear pricing, the optimal insurance contract is an excess-of-loss contract with a deductible. In the standard setting where each loss-causing event generates at most one claim, a per-claim deductible coincides with a per-event deductible, and per-claim excess-of-loss coverage is optimal because it insures the largest losses.

If a single event may generate multiple claims, this equivalence breaks down. The relevant loss to consider is then the aggregate event loss rather than the individual claim. In that setting, the appropriate benchmark for optimal design is a \textit{per-event} excess-of-loss contract that pools all claims from the same underlying cause and covers the tail of the pooled loss.

In our comparisons, we abstract from this distinction and treat $N$ as the relevant count of losses. This simplification does not invalidate the welfare comparison; if anything, when per-claim deductibles are no longer optimal, it becomes easier for parametric insurance to dominate a (now) suboptimal deductible benchmark, all else equal.
\end{remark}

\begin{remark}[Basis risk and richer parametric designs]\label{R:basisrisk}
\citet[in a mean-variance setting similar to ours]{HuSh24} compare indemnity insurance with index insurance. They show that improving the index reduces basis risk and brings the parametric cover closer to indemnity in expected-utility terms. Their results, therefore, support an important implication for our analysis: moving beyond the constant benefit $k$ and reducing basis risk through index triggers or richer indemnity functions can make the welfare case for parametric insurance even more compelling. For example, \citet{MaSc26} characterize the basis-risk-optimal parametric
payment scheme as a conditional expectile of the true loss given that the index triggers, providing an explicit construction of the richer designs referenced above.

In practice, basis risk can arise from misalignment in both $Y_i$ and $N$, because the trigger is typically defined using an auxiliary indicator (e.g., \ wind speed) rather than the presence and magnitude of actual loss. That said, FloodFlash requires proof of (any) actual damage before payment in the US (a requirement that does not exist in the UK), which materially reduces basis risk on $N$.

Our optimization below focuses on the utility loss from basis risk in $Y_i$. A more general setup could model two dependent processes $N_{\text{true}}$ and $N_{\text{trigger}}$, with $\Ben{d}{N,Y_1,\dots,Y_N,d}$ formulated on $N_{\text{true}}$ and $\Ben{p}{N,k}$ on $N_{\text{trigger}}$. When $(N_{\text{true}},N_{\text{trigger}})$ form dependent compound Poisson processes, the claim counts decompose into common and idiosyncratic components, and calculations remain straightforward.
\end{remark}

\subsection{Premia}

For the traditional indemnity benefit $\Ben{d}{N,Y_1,\dots,Y_N,d}$ we define the premium under the expectation principle as
\begin{align}
\label{Pd}
\prem{d}{d}{\theta_d,\gamma_d}
  &= (1+\theta_d)\bigl(E[\Ben{d}{N,Y_1,\dots,Y_N,d}]+\gamma_d\bigr)
   = (1+\theta_d)\bigl(E[(Y_i-d)_+]E[N]+\gamma_d\bigr),
\end{align}
where $\theta_d$ is a loading factor and $\gamma_d$ captures additional expenses not directly proportional to losses (including the ``unallocated loss adjustment expenses'' (ULAE) and other overheads).

For the parametric benefit $\Ben{p}{N,k}$ we analogously set
\begin{align}
\label{Pp}
\prem{p}{k}{\theta_p,\gamma_p}
  &= (1+\theta_p)\bigl(E[\Ben{p}{N,k}]+\gamma_p\bigr)
   = (1+\theta_p)\bigl(kE[N]+\gamma_p\bigr),
\end{align}
with loading $\theta_p$ and fixed-cost component $\gamma_p$ for ULAE. We treat the fixed-cost component as requiring risk capital. Consequently, it is included in the expected-value premium principle and therefore also risk-loaded. Under this interpretation, for each premium the resulting minimum premium $(1+\theta)\gamma$ reflects the insurer’s cost-of-risk requirements even in low-loss regimes.

The loading is intended to compensate the insurer for bearing risk beyond the expected claim cost. In practice, one expects $\theta_d>\theta_p$. Intuitively, the indemnity contract $\Ben{d}{\cdot}$ covers the right tail of severity outcomes (losses beyond $d$), precisely the component that matters most for a risk-averse insured and underlies the classical optimality of deductibles. From the insurer’s perspective, however, tail exposure increases required capital: right-tail quantiles of per-policy benefits are higher under $\Ben{d}{\cdot}$ than under $\Ben{p}{\cdot}$, so a higher loading is economically justified\footnote{\citet{JuJu26} document empirically that, for US homeowner's insurance, the median ratio of expected loss to premium is approximately $28\%$, indicating that premiums substantially exceed expected losses --- consistent with the loadings $(1+\theta_d)$ in our framework taking values materially above unity.}.

Similarly, one typically expects $\gamma_d>\gamma_p$ for three related reasons. First, loss adjustment for indemnity insurance is costlier: assessing and settling $\Ben{d}{\cdot}$ requires measuring actual damages, and catastrophe environments exacerbate these costs because claims are numerous, access can be difficult, and disputes are more likely. Second, post-disaster reconstruction costs often exceed pre-loss insured values due to demand surges for labor and materials; some markets reflect this in policies that pay up to (say) $125\%$ of insured value after major events. Under expectation pricing, an up-scaling of $Y_i$ increases the expected indemnity cost and can therefore be represented as an additive component in the premium that we absorb into $\gamma_d$.\footnote{If pricing involved explicit variance loading or risk measures, the treatment would need to distinguish cost-level changes from risk-level changes, and $\gamma$ would no longer be a sufficient reduced-form summary.} Third, indemnity claims typically take a long time to settle; even when delays translate into direct costs (e.g.\ temporary accommodation) captured by the first component, they also justify an additional frictional penalty reflected by $\gamma_d$.

In \eqref{Pd}--\eqref{Pp} the fixed-cost terms $\gamma_\cdot$ do not multiply $E[N]$. While the first two components above are plausibly proportional to the number of events, the third is not. A fully disaggregated model would therefore introduce multiple fixed-cost parameters. We adopt the reduced-form specification for two reasons: it keeps the analysis transparent, and it allows $\gamma_d$ (especially when $\gamma_p\equiv 0$) to serve as a single quantitative proxy for operational frictions that differentiate indemnity from parametric insurance. Moreover, interpreting $\gamma$ as a \textit{per-period} penalty (rather than per-event) is often more natural in household contexts. A per-event formulation can be recovered by rescaling by $1/E[N]=1/\lambda$ if desired.\footnote{Equivalently, one may interpret $\gamma$ as already incorporating the relevant scaling given the typical event frequency in the market considered.}

Note that conceptually the spread in $\gamma$'s pertain to differences in expected loss, whereas the spread in $\theta$'s pertain to risk premium (the cost of protecting the insurer).

\subsection{Decision making criteria}

The agent evaluates insurance choices using a mean--variance objective
\begin{align}
  \MV{.}{W}{\beta} = E[W] - \beta\cdot \Var(W),
  \qquad \beta>0,
\end{align}
where terminal wealth is
\begin{align}
  W = w_0 - \prem{.}{\cdot}{\cdot} - S + \Ben{.}{N,Y_1,\dots,Y_N,.},
\end{align}
and $w_0$ denotes initial wealth.

A technical feature of mean--variance preferences is that the variance term has units of currency squared, whereas the expectation has units of currency. A common normalization is therefore to scale risk aversion by wealth and set
\begin{equation}
  \beta=\frac{1}{w_0},
\end{equation}
which ensures that both terms in the objective are comparable and aligns with interpreting variance as a relative dispersion measure. In static optimization, this normalization can always be absorbed into the calibration of $\beta$ in numerical work. In dynamic settings, the normalization becomes more delicate; examples from mathematical finance include \citet{JZ2008}, \citet{XZ2013}, and \citet{BMZ2014}, who scale the variance by wealth similarly to us.

Each insurance design has a single contract parameter to be chosen. We define the optimal deductible and optimal parametric payment by
\begin{align}
  d^* = \arg\max_{d}\ \MV{d}{W}{d,\beta,\theta_d,\gamma_d},
  \qquad
  k^* = \arg\max_{k}\ \MV{p}{W}{k,\beta,\theta_p,\gamma_p}.
\end{align}

Closed-form expressions for expected payouts, second moments, and the resulting mean--variance objectives under censored exponential severity and Poisson frequency are collected in Appendix~\ref{App:MV_censored_exp}.

\subsection{Research question: indemnity vs parametric insurance}

Under the expectation premium principle, where the premium loading is proportional to expected loss, classical results imply that the deductible structure $\Ben{d}{N,Y_1,\dots,Y_N,d}$ is optimal under expected-utility maximization; see \citet{Bor69,Arr74}. Our goal is to revisit this benchmark in settings motivated by catastrophic household risks, where the traditional paradigm is strained by affordability, availability, and operational frictions.

The key economic motivation is that indemnity insurance is typically priced and designed to cover (or at least credibly underwrite) the full underlying loss distribution. When full cover becomes unaffordable, the main actuarial lever to reduce the premium is to increase the deductible—often to levels that are difficult to interpret as meaningful household protection (a mechanism we will illustrate numerically). By contrast, parametric insurance is naturally scalable: the insured can purchase a modest layer of per-event protection $k$ at a correspondingly modest premium, and the benefit can be paid quickly upon verification of the trigger, without requiring the household to incur a huge out-of-pocket loss before receiving any payment. Unlike much of the existing literature on index insurance, which takes partial coverage as given, we explicitly compare how partial coverage is achieved under indemnity versus parametric designs, and show that the two scaling mechanisms have fundamentally different welfare implications.

This paper, therefore, investigates whether and when parametric insurance can be optimal (or welfare-improving) once realistic frictions and constraints are introduced. Concretely, we allow for (i) differences in operational and settlement frictions summarized by fixed-cost terms $\gamma_d$ and $\gamma_p$; (ii) potentially different loadings $\theta_d$ and $\theta_p$ reflecting different capital requirements and severity-risk exposures; and (iii) explicit premium budget constraints that may bind for households in high-risk regions. Finally, the question is timely in light of emerging policy discussions around parametric home insurance for catastrophic perils. For example, in some jurisdictions \citep[e.g.\ Australia; see][]{JaMeMa24} there are calls for the introduction of parametric home insurance for catastrophic risks such as floods, cyclones, or bushfires. The subsequent sections analyze these mechanisms theoretically and numerically.

\begin{remark}[Discussion of the mean--variance choice]\label{R:MV_choice}
We adopt a mean--variance objective for tractability and because it delivers closed-form comparative statics without requiring the specification of an arbitrary utility function. A familiar concern with mean--variance preferences is that they penalise symmetric dispersion  and may therefore understate the value of protection against severe outcomes. However, this concern is muted in our setting because the loss support is bounded on both sides (property insurance pays only up to the sum insured; see also the next section and Remark \ref{R:severity_scope}), so the distribution has no unbounded tail to misrepresent. In addition, we deliberately adopt the most conservative parametric design: a constant-per-event payment, which embeds the maximum possible basis risk. Richer designs (piecewise-constant, indexed, or capped payouts; see Section~\ref{sec:framework} and Remark~\ref{R:basisrisk}) would only strengthen the welfare case for parametric insurance. Any welfare advantage established under our specification should therefore be read as a lower bound, rather than an exact reading. If parametric insurance dominates in regions of our setting, it is even more likely to do so under more realistic specifications.
\end{remark}

\section{Explicit results in the compound Poisson case} \label{sec:model}

In this section, we make specific assumptions about the distributions of $N$ and $Y_i$ to obtain explicit results for the optimal strategy parameters, as well as the corresponding premia and expected utility levels.

\subsection{Dual relationship between $d^*$ and $k^*$ in the compound Poisson case} \label{E_identity}

We now assume that the claim count $N$ is Poisson distributed with mean $\lambda$, so that the aggregate loss $S = \sum_{i=1}^N Y_i$ follows a compound Poisson distribution. Under this assumption (which we will further discuss in Section \ref{S:modelingchoices}), and provided that preferences are 
described by the mean–variance objective, and insurance premiums are calculated according to the expectation principle, we obtain explicit and straightforward expressions for both the optimal deductible $d^{*}$ and the optimal parametric per–event cover $k^{*}$. We further assume that $\theta_d=\theta_p$.

The key mathematical property underlying these results is the equi-dispersion identity for Poisson random variables,
\begin{align*}
  \Var(N) = E[N] = \lambda,    
\end{align*}
which implies that the variance of a random sum of i.i.d.\ terms reduces to
\begin{align*}
\Var\!\left(\sum_{i=1}^N Y_i\right)
     = \lambda\, E[Y_i^{2}].
\end{align*}
This linear structure ensures that, under the mean–variance objective, the variance component of terminal wealth does not depend on nonlinear interactions between the distribution of $N$ and the deductible or per–event cover.

In Appendix \ref{App:Duality}, we show that, under the combined assumptions of 
(i) the mean–variance objective, 
(ii) the Poisson claim number model, and 
(iii) the expectation principle for premiums, 
the first-order conditions for the two optimization problems reduce to
\begin{align} \label{E_dstar_kstar}
d^{*} = \frac{\theta}{2\beta},
\qquad
k^{*} = E[Y_i] - \frac{\theta}{2\beta},
\end{align}
when the loading factors coincide, i.e., $\theta_d = \theta_p = \theta $. Importantly, these expressions do not depend on the distributional form of the claim severity, as long as it has a finite second moment.

A direct implication is the following duality identity:
\begin{equation} 
E[Y_i]
  = d^{*} + k^{*}.
  \label{eq:duality}
\end{equation}
This relation has a natural economic interpretation: $d^{*}$ is the optimal amount of protection removed from full insurance, whereas $k^{*}$ is the optimal amount of protection added from no cover. The two adjustments sum precisely to the mean insured loss.

It is crucial to emphasize that the duality is not universal. As detailed in Appendix \ref{App:Duality}, the identity relies essentially on all three 
assumptions listed above. If one replaces the expectation principle by any premium calculation involving variance loading or risk measures, or if $N$ does not satisfy 
$\Var(N) = E[N]$, the optimality conditions no longer reduce to the simple linear forms above, and the duality identity fails. We also stress that the dualty relation is based on interior solutions, i.e., no budget or risk constraints.

\begin{remark}[Why fixed costs $\gamma_\cdot$ do not appear in the FOC]
    Interestingly $\gamma_\cdot$ does not feature in \eqref{E_dstar_kstar}. This is because it affects expected wealth but not the insured's marginal risk–return trade-off. Under mean–variance preferences and expectation-principle pricing, optimal contract parameters are determined entirely by marginal premium loadings and variance reduction, so additive premium components vanish from the first-order conditions.
\end{remark}

\subsection{Optimal results under censored exponential losses}

Assume now that $Y_i$ is a censored exponential random variable with mixed density
\begin{equation} \label{E_losses}
f_{Y_i}(y)=
\begin{cases}
\nu e^{-\nu y}, & y \in [0,L);\\
e^{-\nu L},     & y = L;\\
0,              & \text{otherwise}.
\end{cases}
\end{equation}
This structure is justified by property insurance, which typically has a limited sum at risk $L$ (the value of the property). Furthermore, there will typically be a positive probability of complete write-off (here $e^{-\nu L}$). Of course, the exponential assumption here simplifies some calculations, but it is not unreasonable in an insurance context, and we don't expect that the use of a more sophisticated distribution would materially alter our conclusions.

\begin{remark}[Loss limit $L$ vs policy limit $M$]
The limit $L$ is formulated on the loss. A \textit{policy limit} $M$ would be applied on the benefit $\Ben{.}{N,Y_1,\dots,Y_N,.}$ such that $Y_i-d$ (the loss \textit{net of the deductible)} would be capped at $M$. Calculations can easily be extended to this case, but do not lead to materially different conclusions and hence have been omitted here.
\end{remark}

\begin{remark}[Degenerate severity in the limit $\nu \to 0$]
    If $\nu \longrightarrow 0$ for fixed $L$ then the distribution in \eqref{E_losses} becomes point-mass ($L$ with probability 1). In this case we retain $E[Y_i]=L=d^* +k^*$, however $d^* \neq k^*$ in general. Both policies become identical, with payout $k^* = L-d^*$ for each event.
\end{remark}

Let $Y_i$ have the density function of equation \eqref{E_losses} so that
\begin{equation}
dF_{ Y_i}(y) = \nu e^{-\nu y}\,dy + e^{-\nu L}\,\delta_L(dy).
\end{equation}

From Appendix~\ref{App:MV_censored_exp}, we have that the expected mean-variance of the indemnity cover is
\begin{equation} \MV{d}{W}{d,\beta,\theta_d,\gamma_d}
 = w_0 - \prem{d}{d}{\theta_d,\gamma_d}
 - \lambda E[Y_i]
 + \lambda E[(Y_i - d)_+]
 - \beta \lambda \frac{2}{\nu^2}\big[1 - e^{-\nu d}(1 + \nu d)\big].
\end{equation}

Differentiation yields
\begin{equation}
\frac{\partial \text{MV}^{(d)}}{\partial d}
 = \lambda e^{-\nu d}(\theta_d - 2\beta d), 
\qquad
\frac{\partial^2 \text{MV}^{(d)}}{\partial d^2}
 = \lambda e^{-\nu d}(-\nu\theta_d + 2\nu\beta d - 2\beta),
\end{equation}
so that
\begin{equation}
d^{*} = \frac{\theta_d}{2\beta}, 
\qquad
\frac{\partial^2 \text{MV}^{(d)}}{\partial d^2}\Big|_{d^{*}} = -2 \beta \lambda e^{-\nu d^{*}}< 0.
\end{equation}

On the other hand, the expected mean-variance of the parametric cover is
\begin{equation} 
\MV{p}{W}{k,\beta,\theta_p,\gamma_p}
 = w_0 - \prem{p}{k}{\theta_p,\gamma_p}
 - \lambda E[Y_i] + \lambda k
 - \beta\lambda\big(\mathrm{Var}(Y_i) + (E[Y_i] - k)^2\big);
\end{equation}
see Appendix~\ref{App:MV_censored_exp}. Differentiation gives
\begin{equation}
\frac{\partial \text{MV}^{(p)}}{\partial k}
 = \lambda(-\theta_p + 2\beta(E[Y_i] - k)), 
\qquad
\frac{\partial^2 \text{MV}^{(p)}}{\partial k^2}
 = -2\beta\lambda < 0,
\end{equation}
so
\begin{equation}
k^{*} = E[Y_i] - \frac{\theta_p}{2\beta}.
\end{equation}

\begin{remark}[Scope of the severity model]\label{R:severity_scope}
The censored exponential specification is tailored to \textit{household property insurance}, where the sum insured $L$ (the value of the dwelling) provides a natural and contractually enforced upper bound on the insurer's severity exposure. In this setting, the cap is a feature of the underlying contract, not a modeling approximation: insured losses above $L$ are by construction not transferable to the insurer. The specification is also conservative in that a substantial probability mass at $L$ (complete write-off) concentrates loss density at large outcomes\footnote{Note this is consistent with the empirically observed right-skewness of household property losses in \citet{JuJu26}.}, which favors excess-of-loss indemnity cover over a constant-per-event parametric payment. Establishing that parametric cover can dominate in this environment therefore strengthens the result.
\end{remark}

\subsection{Interpretation: tail coverage versus first-dollar coverage}
\label{sec:tail_vs_first_dollar}

The identity \eqref{eq:duality} admits a clear economic interpretation of how indemnity and parametric contracts allocate protection across the loss distribution. Under the optimal excess-of-loss indemnity contract, insurance coverage applies only to losses exceeding the deductible $d^*$: indemnity insurance therefore protects the \textit{tail} of the loss distribution. By contrast, the parametric contract pays a fixed amount $k^*$ whenever an insured event occurs, independently of its severity. Parametric insurance thus provides \textit{first-dollar} protection.

This distinction is largely irrelevant when contracts are unconstrained and risk is low. However, it becomes decisive once affordability or budget constraints are introduced. Scaling down a parametric contract by reducing $k$ proportionally reduces first-dollar protection while preserving its qualitative role: the insured continues to receive immediate liquidity whenever an event occurs. Scaling down an indemnity contract, by contrast, requires increasing the deductible $d$, which removes coverage precisely where losses are most likely to materialize. As a result, partial indemnity insurance rapidly becomes economically ineffective: very high deductibles leave households self-insuring most realizations, even though coverage remains formally available.

This asymmetry explains why budget constraints affect the two designs so differently. Parametric insurance remains scalable and meaningful at low budgets because it adds protection ``from the ground up''. Indemnity insurance, on the other hand, can only become affordable by covering ever thinner slices of the tail of the loss distribution. In high-risk environments, where premiums are dominated by fixed costs and capital loadings, this mechanism renders partial indemnity impractical well before it becomes formally infeasible.

Finally, the provision of first-dollar protection is often viewed as undesirable in traditional insurance settings because of moral hazard concerns. In the context of disaster risk, however, this argument carries much less force. Households have no control over the occurrence of floods, storms, or wildfires, and therefore cannot influence the triggering of parametric payouts. Moreover, because the parametric payment is fixed, insured households retain strong---and arguably stronger---incentives to mitigate loss severity, as any reduction in damage directly benefits them. In this setting, the absence of a deductible is therefore not a drawback of parametric insurance, but a feature that enhances its effectiveness when only partial coverage is affordable.

\subsection{Robustness and modeling choices} \label{S:modelingchoices}

The modeling choices adopted here are deliberately conservative. The parametric contract is specified as a constant per-event payment, which maximizes basis risk and therefore places parametric insurance at a theoretical disadvantage relative to indemnity insurance. Allowing for more flexible parametric designs—such as piecewise-constant payouts or index-linked triggers—would mechanically reduce basis risk. Therefore, it is unlikely to overturn our conclusions.

Similarly, the censored exponential severity distribution is neither light- nor heavy-tailed, and serves primarily to deliver closed-form expressions. In environments with heavier-tailed losses, the capital intensity and fixed costs of indemnity insurance are likely to be even more pronounced, strengthening the relative appeal of parametric
coverage under budget constraints.

The compound Poisson framework adopted here is standard in catastrophe risk modeling \citep[see, e.g.,][]{ChBuRaTrWe06,EkHo14,HiFoMiJo17}, and its memoryless property is not binding in the low-frequency regime that motivates our analysis: at typical annual return periods, within-year clustering of events plays a negligible role in the welfare comparison. Here, the ``memoryless'' is best reframed as (its mathematically equivalent) ``constant hazard rate'' within the coverage period. That being said, two possible extensions of the Poisson assumption are worth noting. First, \textit{across years}, reconstruction choices could affect frequency assumptions; see also Remark \ref{R:resilience}. Second, count overdispersion is easy to implement numerically and does not appear to weaken our results; see also Remark \ref{R:overdispersion}.

\begin{remark}[Allowance for resilience measures]\label{R:resilience}
Across years, reconstruction choices can influence the prevailing intensity: for example, ``Build Back Better'' programs \citep{UNDRR2015,FloodRe2022,EIOPAECB2024} aim to reduce vulnerability and therefore effectively shift the loss intensity downward in subsequent periods. Such dynamics lie outside the static single-period framework adopted here, but they are not precluded by it: our mean--variance comparison applies \textit{period-by-period}, and the relevant pre- and post-event parameters can be re-calibrated to reflect adaptation. A dynamic extension with endogenous adaptation is a natural avenue for future work.
\end{remark}

\section{Numerical illustrations}\label{sec:numerics}

This section illustrates the welfare comparison between indemnity (excess-of-loss) and parametric insurance in a low-frequency, high-severity environment, using the closed-form expressions derived in Sections~\ref{sec:framework}--\ref{sec:model} (and implemented in the companion numerical appendix).
Throughout, we report \textit{mean--variance} (MV) values, with larger values corresponding to higher welfare.

\subsection{Baseline calibration}\label{subsec:calibration}

Consider a household with a house worth \$500,000. Their wealth corresponds to 30\% of the value of the house, that is, \$150,000. The house is built in a flood plain with a 1-in-50-year chance of severe flooding ($\lambda=1/50$). Such an event will happen at least twice in a lifetime (of 80 years) with probability 48\%.

Should such a flood occur, damages to the house follow the distribution in \eqref{E_losses}, with $\nu=1/350,000$. This means $E[Y_i] = \$266,122$, and the probability of a complete write-off is $\Pr[Y_i=L]=24\%$. Importantly, a large amount of loss density is in its tail, which will naturally favor an excess-of-loss cover over a parametric cover; see also Remark \ref{R:severity_scope}.

This set-up corresponds to a realistic flood risk scenario. In Ipswich (Queensland, Australia), a 1-in-50-year event corresponds to a gauge at 18.7m in its CBD \citep{QRA19}. This is a similar level to that observed in the 2011 floods in this region, which saw many homes completely written off. 

\subsection{Unconstrained optima and one-dimensional comparative statics}\label{subsec:1d}

We begin by considering the case $\gamma_p\equiv0$ and $\theta_d=\theta_p\equiv\theta=0.3$. Whence, we compare the expected mean-variance utility of both covers for varying levels of $\gamma_d \ge 0$. In this case,
\begin{equation}
d^*=\$ 22,\!500\quad\text{ and }\quad k^* = \$ 243,\!622.
\end{equation}
Corresponding premiums are (at the baseline $\gamma_d = 0$)
\begin{equation}
\prem{d}{d^*}{\theta,0}=\$ 6,\!353\quad\text{ and } \quad \prem{p}{k^*}{\theta,0} = \$ 6,\!334.
\end{equation}

\begin{figure}[htb]
  \centering
  \captionsetup{font=small}
  
  \begin{minipage}[t]{0.62\textwidth}
    \centering
    \includegraphics[width=\textwidth]{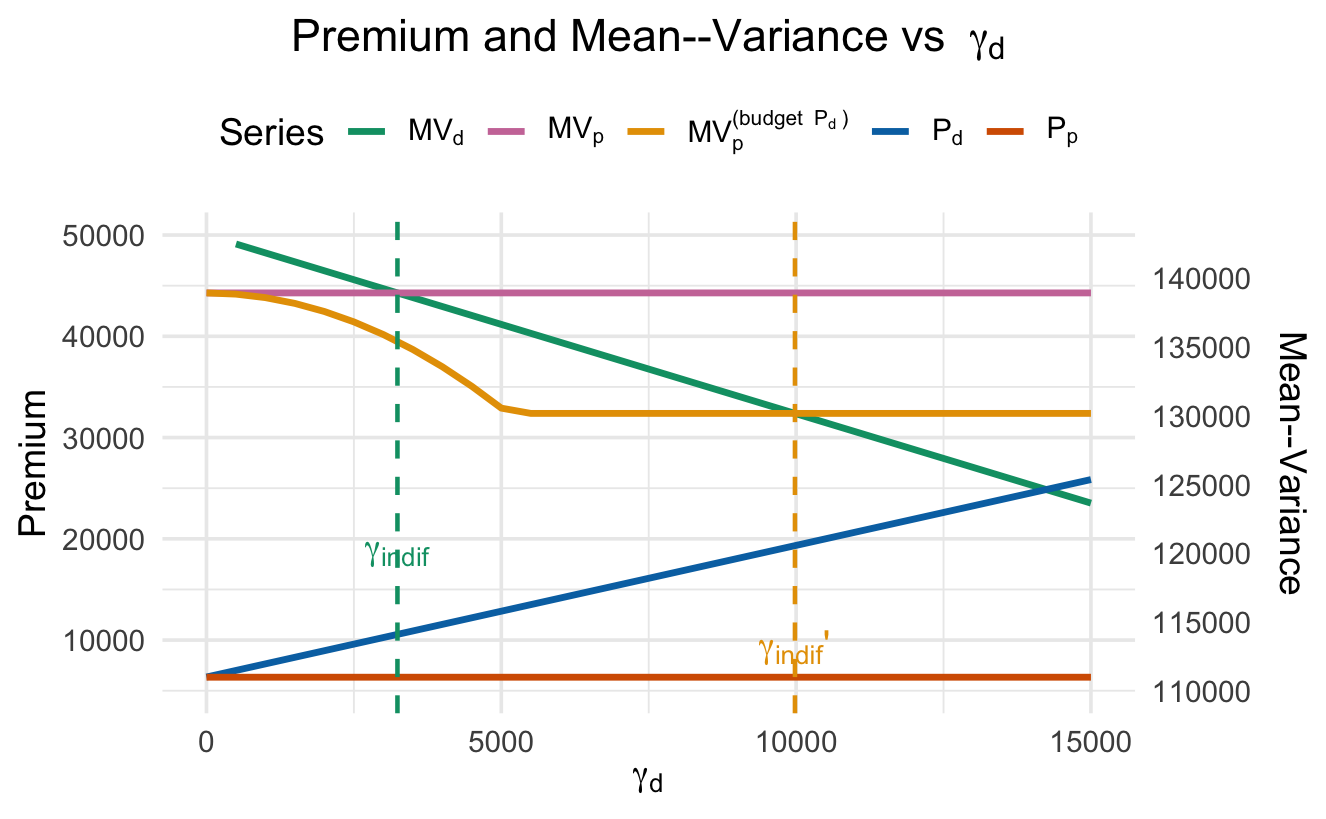}
    
    \smallskip
    {\small (a) One-dimensional comparative statics in $\gamma_d$: premiums and
    mean--variance at unconstrained optima $d^*(\theta)$ and $k^*(\theta)$.}
    
    %\label{fig:gamma_1d}
  \end{minipage}
  \hfill
  \begin{minipage}[t]{0.36\textwidth}
    \centering
    \includegraphics[width=\textwidth]{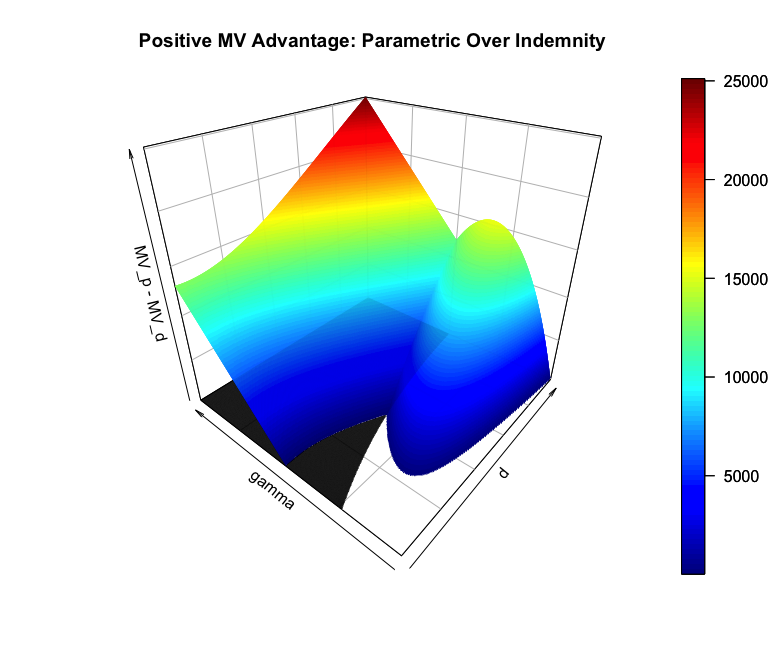}
    
    \smallskip
    {\small (b) Premium-matching surface over $(d,\gamma_d)$: positive part of
    $MV^{(p)}-MV^{(d)}$, with the cap region $\{k=L\}$ highlighted on the floor.}
    
    %\label{fig:surface_d_gamma_premiummatch}
  \end{minipage}

  \caption{Comparison of indemnity and parametric insurance under premium
  matching. Panel (a) shows one-dimensional comparative statics in the
  indemnity fixed cost $\gamma_d$, while panel (b) extends the comparison to
  the two-dimensional $(d,\gamma_d)$ space and highlights the region in which
  parametric insurance yields higher mean--variance utility.}
  \label{fig:gamma_and_surface}
\end{figure}

Figure \ref{fig:gamma_and_surface}(a) shows levels of premium as per \eqref{Pd}-\eqref{Pp} at the bottom (coordinates on the left). The premium for the indemnity insurance increases linearly with $\gamma_d$ with slope $(1+\theta)$, whereas the premium for parametric insurance is not affected by $\gamma_d$ and remains constant. The expected mean-variance functions are shown in green and purple, respectively. The level of $\gamma_d$ that makes an agent indifferent between either coverage is \$3,239. If one forces the agent to spend as much on the parametric cover as they spend on the indemnity cover (increasing $k$ beyond its optimal level), the corresponding expected mean-variance (in orange) decreases until $k=L$ and no additional cover can be purchased, after which it stays constant. Under such a scheme, the level of $\gamma_d$ that makes the agent indifferent between the two covers is larger and sits at \$9,980.

The idea of varying $k$ to match premia is generalized in Figure \ref{fig:gamma_and_surface}(b). For each grid point $(d,\gamma_d)$ we compute the indemnity premium $\pi^{(d)}(d;\theta_d,\gamma_d)$ and choose the parametric payment
\begin{equation}\label{eq:k-premium-match}
  k(d,\gamma_d)
  = \min\!\left\{
      \frac{\prem{d}{d}{\theta_d,\gamma_d}/(1+\theta_p)-\gamma_p}{\lambda},
      \,L
    \right\},
\end{equation}
so that the parametric premium matches the indemnity premium whenever the implied $k$ lies below $L$. The MV difference surface then plots
$MV^{(p)}(k(d,\gamma_d);\theta_p,\gamma_p)-MV^{(d)}(d;\theta_d,\gamma_d)$, truncated at zero so that only regions where parametric is strictly better remain visible.

When $k(d,\gamma_d)<L$, the mapping in \eqref{eq:k-premium-match} is smooth and the MV difference varies smoothly in $(d,\gamma_d)$. When the implied payment exceeds $L$, the constraint $k\le L$ binds, creating a \textit{corner solution}: parametric protection cannot increase further even if the indemnity premium continues to increase. In the corresponding surface plots, we explicitly mark the set $\{(d,\gamma_d):k(d,\gamma_d)=L\}$ (the \textit{cap region}) in black on the ``floor''. The black region, therefore, identifies \textit{where the premium-matching rule hits the parametric maximum payment}, not where the MV difference is negative.

One can see that the indemnity cover is indeed always optimal for
$\gamma=0$, but it becomes suboptimal for tiny $\gamma$'s already when
$d$ is very suboptimal (large).

An analogous construction applies over $(\theta_d,\gamma_d)$, where indemnity is evaluated at $d^*(\theta_d)$ and parametric is premium-matched to $\prem{d}{d^*(\theta_d)}{\theta_d,\gamma_d}$, yielding a surface in the fundamental pricing parameter $\theta_d>\theta_p=0.2$ (cost of coverage) rather than contract parameters. Here $\gamma_d=\gamma_p=0$.

\begin{figure}[htb]
  \centering
  \captionsetup{font=small}

  \begin{minipage}[t]{0.62\textwidth}
    \centering
    \includegraphics[width=\textwidth]{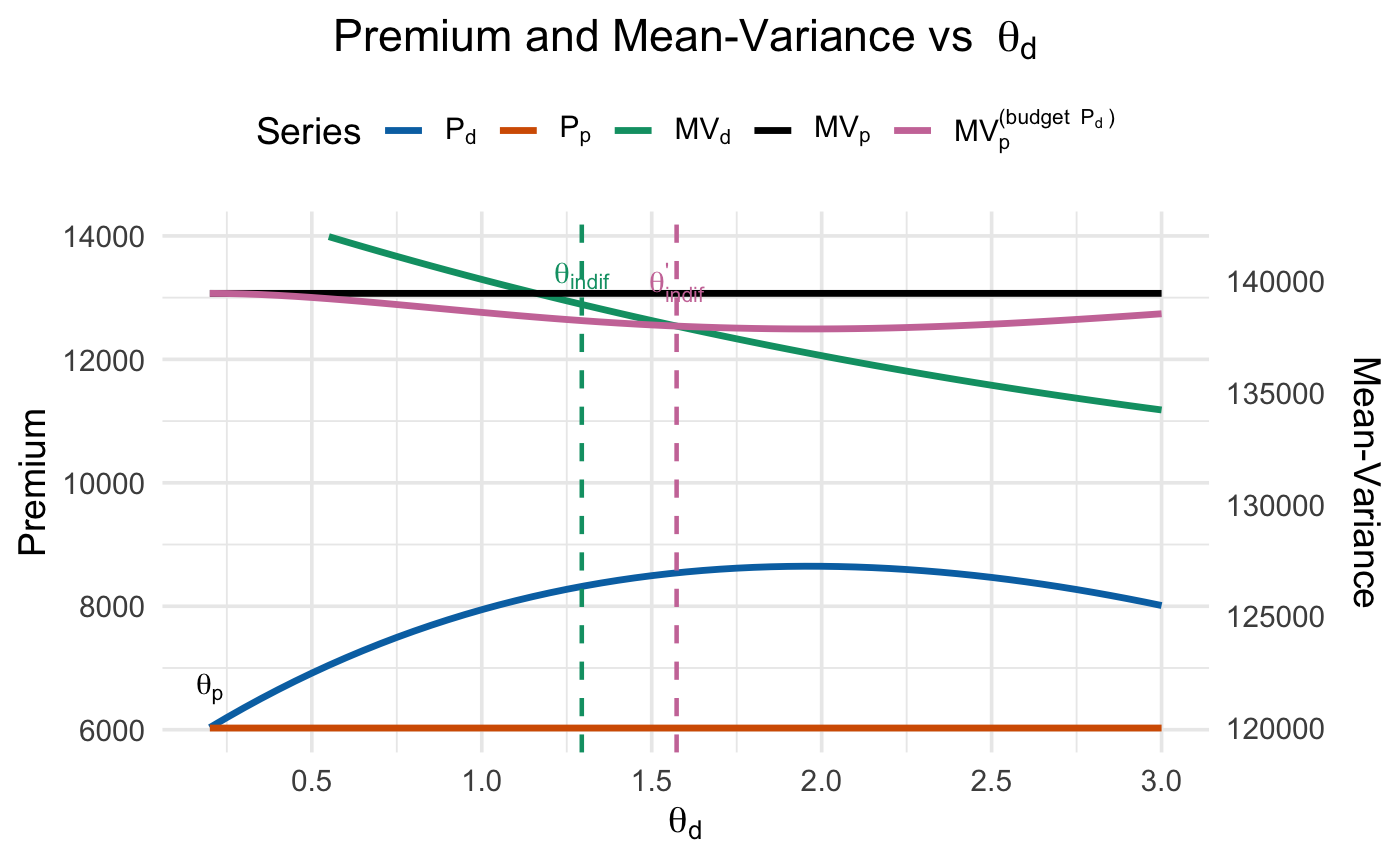}
    
    \smallskip
    {\small (a) One-dimensional comparative statics in the indemnity loading
    $\theta_d$: premiums and mean--variance at unconstrained optima
    $d^*(\theta_d)$ and $k^*(\theta_p)$, together with the premium-matched
    parametric value $MV_p^{(\text{budget }P_d)}$.}
    
    %\label{fig:theta_1d}
  \end{minipage}
  \hfill
  \begin{minipage}[t]{0.36\textwidth}
    \centering
    \includegraphics[width=\textwidth]{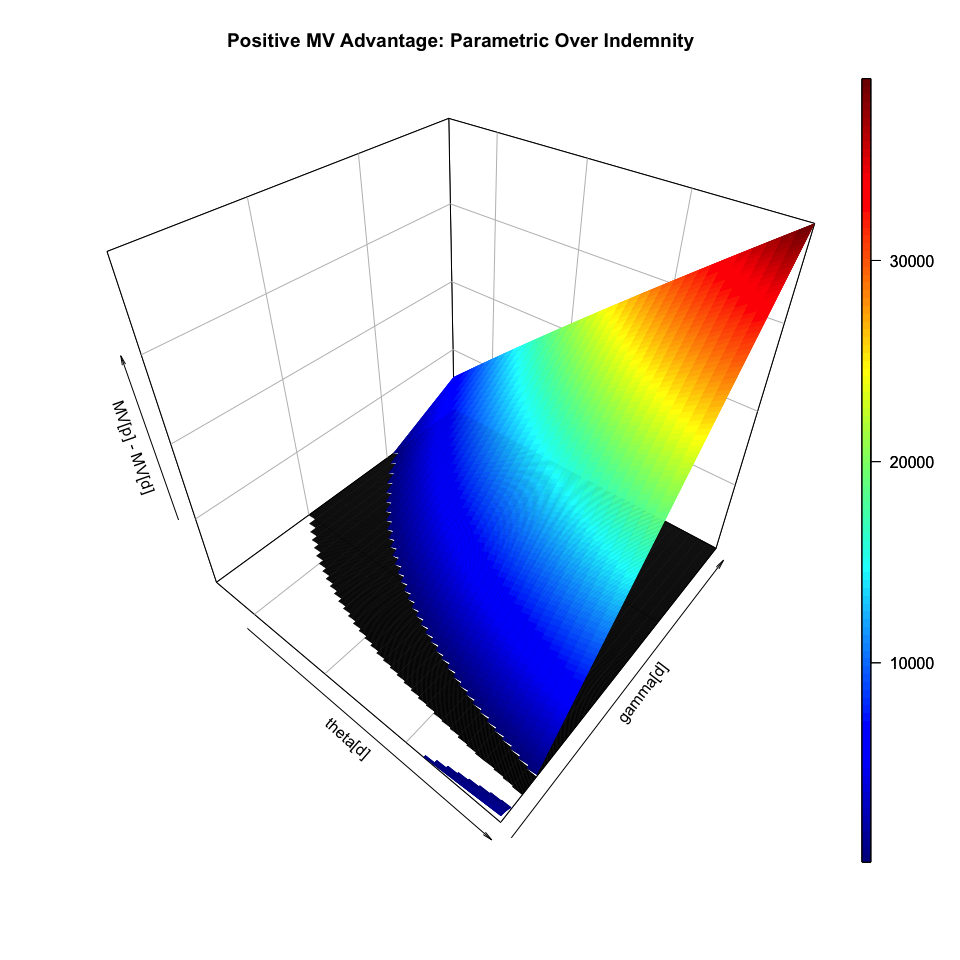}
    
    \smallskip
    {\small (b) Premium-matching surface over $(\theta_d,\gamma_d)$:
    The parametric contract is calibrated to match the premium of the
    optimal indemnity contract $d^*(\theta_d)$ at each point.
    The surface shows the positive part of $MV^{(p)}-MV^{(d)}$.}
    
    %\label{fig:surface_theta_gamma_premiummatch}
  \end{minipage}

  \caption{Premium matching in the space of indemnity pricing parameters.
  Panel (a) shows how premiums and mean--variance utilities vary with the
  indemnity loading $\theta_d$ in one dimension. Panel (b) extends the
  comparison to the two-dimensional $(\theta_d,\gamma_d)$ space and
  identifies regions in which the premium-matched parametric contract
  delivers higher mean--variance utility than the optimal indemnity contract.}
  \label{fig:theta_and_surface_premiummatch}
\end{figure}

The interpretation of Figure \ref{fig:theta_and_surface_premiummatch} is analogous to that of Figure \ref{fig:gamma_and_surface}. Here, the $\theta_d$ that leads to indifference between both covers is $1.29$ without premium matching, and $1.57$ with premium matching\footnote{To assess the reasonableness of those quantities, recall footnote 4. A $\theta_d$ corresponding to a loss ratio of 0.28 \citep[from][]{JuJu26} is roughly 2.6, well above the indifference quantities calculated here.}. It can be shown that such a root exists and is unique on the range of sensible values for $\theta$ (those that lead to $d^* \le L$).

Figure~\ref{fig:theta_and_surface_premiummatch} illustrates how the welfare
comparison between indemnity and parametric insurance evolves as the
indemnity pricing parameters change. While panel~(a) highlights the
non-monotonic behavior of premium-matched parametric utility as a function
of $\theta_d$, panel~(b) shows that this mechanism persists once fixed costs
$\gamma_d$ are introduced, and identifies the region in which parametric
insurance dominates under equal-premium constraints.

\begin{figure}[htb]
  \centering
  \captionsetup{font=small}

  \begin{minipage}[t]{0.62\textwidth}
    \centering
    \includegraphics[width=\textwidth]{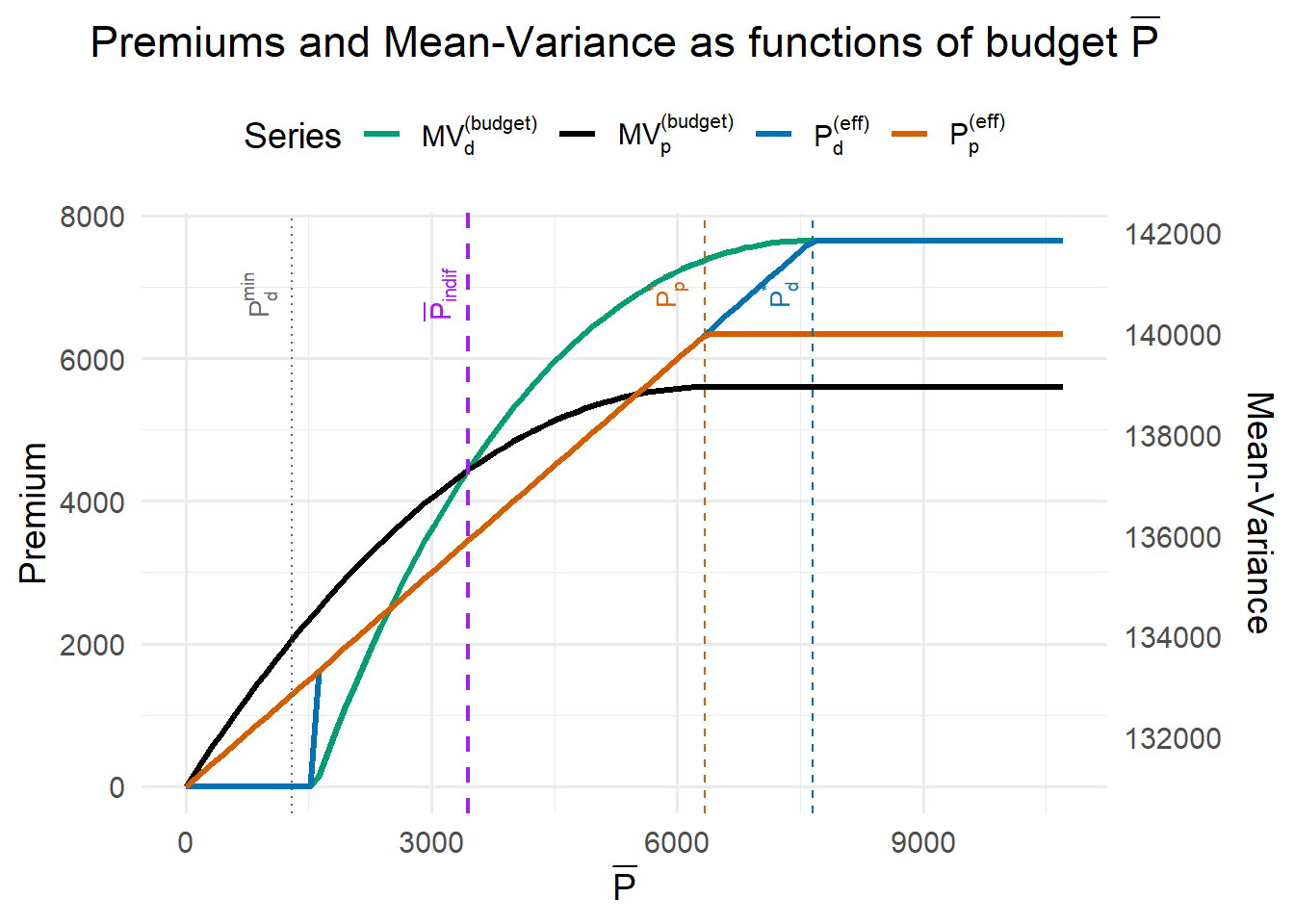}
    
    \smallskip
    {\small (a) Fixed $\gamma_d$: effective premiums (left axis) and
    budget-constrained mean--variance values (right axis) as functions of the
    available premium budget $\bar{P}$. Vertical lines indicate
    $P_d^{\min}$, $P_d^*$, $P_p^*$, and the indifference budget
    $\bar{P}_{\mathrm{indif}}$ (when it exists).}
    
    %\label{fig:budget_1d_dual}
  \end{minipage}
  \hfill
  \begin{minipage}[t]{0.36\textwidth}
    \centering
    \includegraphics[width=\textwidth]{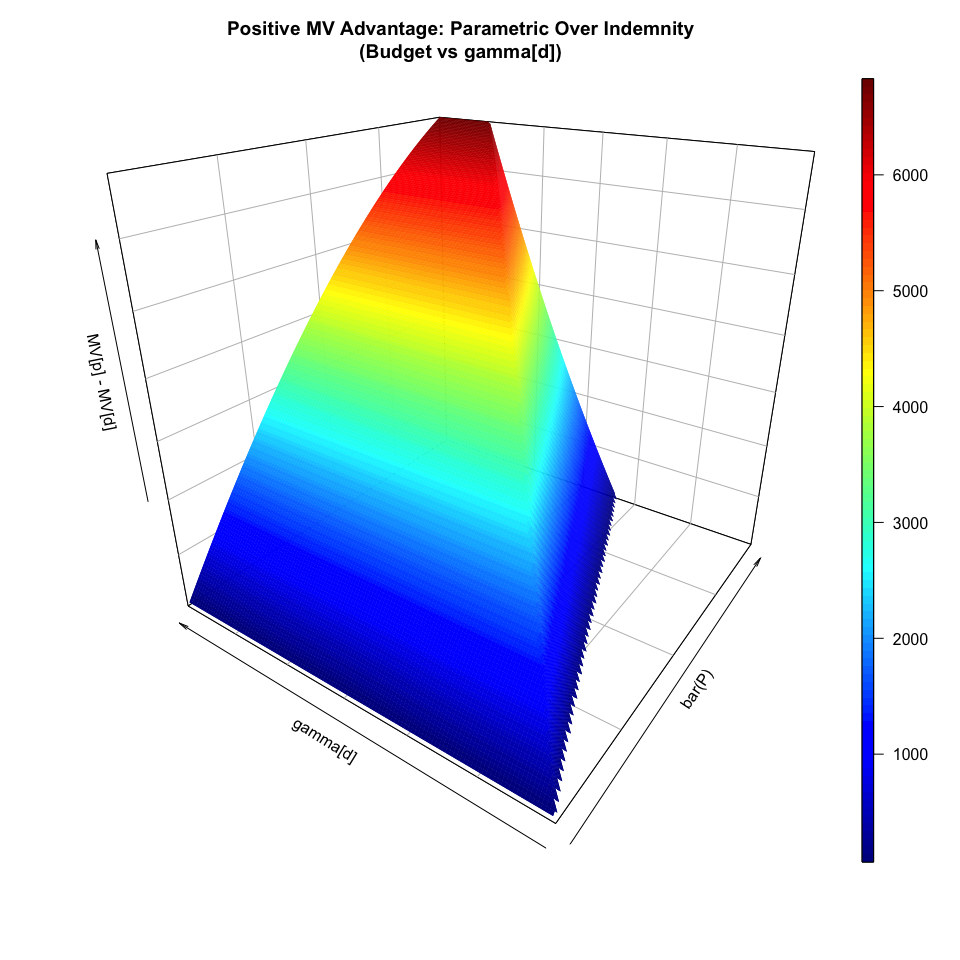}
    
    \smallskip
    {\small (b) Positive region of
    $\Delta MV(\bar{P},\gamma_d)$ in \eqref{eq:surf-def}:
    parameter combinations for which the parametric contract yields strictly
    higher mean--variance utility than the best available alternative
    (indemnity within budget or no insurance).}
    
    %\label{fig:budget_gamma_surface}
  \end{minipage}

  \caption{Budget-constrained comparison of indemnity and parametric insurance.
  Panel (a) illustrates how optimal contract choice and welfare evolve as the
  available premium budget $\bar{P}$ increases for a fixed cost
  $\gamma_d$. Panel (b) extends the analysis to the two-dimensional
  $(\bar{P},\gamma_d)$ space and highlights the region in which parametric
  insurance strictly dominates the best budget-feasible alternative.}
  \label{fig:budget_combined}
\end{figure}

\begin{remark}[Impact of overdispersion on the results]\label{R:overdispersion}
We refer back to the choice of Poisson frequency, as discussed in Section \ref{S:modelingchoices}. A negative-binomial (overdispersed Poisson) sensitivity analysis (not reported for brevity) suggests that, although both optimal contract parameters $d^{*}$ and $k^{*}$ move substantially with count overdispersion, the indifference threshold $\gamma_{d}$ between parametric and indemnity insurance is essentially invariant in absolute terms and \textit{decreases} as a share of premium under higher count volatility. Overdispersion therefore reinforces rather than weakens the case for parametric coverage.
\end{remark}

\subsection{Budget-constrained choice and two-dimensional budget surfaces}\label{subsec:budget}

We now adopt the comparison most relevant for affordability: an agent faces an exogenous premium budget $\bar{P}$ and chooses the best contract they can buy \textit{within} that budget. Critically, the agent can always opt out and remain uninsured. This formulation captures the practical asymmetry: under indemnity pricing with fixed costs, low budgets may translate into either infeasibility or an effectively useless deductible layer, whereas the same budget translates mechanically into a positive parametric payment as long as $\bar{P}>(1+\theta_p)\gamma_p$. This pattern reflects the fact that parametric insurance provides first-dollar liquidity even at low budgets, whereas indemnity insurance becomes affordable only by pushing coverage into the extreme tail.

For each $\bar{P}$, the parametric choice solves
\begin{equation}\label{eq:param-budget}
  MV^{(p)}_{\mathrm{bud}}(\bar{P})
  =
  \max\Bigl\{
      MV^{(0)},
      \ \max_{0\le k\le L:\ \mathcal{P}^{(p)}(k)\le \bar{P}}
      MV^{(p)}(k;\theta_p,\gamma_p)
    \Bigr\}.
\end{equation}
Similarly, given $(\gamma_d,\theta_d)$, the indemnity choice solves
\begin{equation}\label{eq:indem-budget}
  MV^{(d)}_{\mathrm{bud}}(\bar{P},\gamma_d)
  =
  \max\Bigl\{
      MV^{(0)},
      \ \max_{0\le d\le L:\ \mathcal{P}^{(d)}(d)\le \bar{P}}
      MV^{(d)}(d;\theta_d,\gamma_d)
    \Bigr\}.
\end{equation}
Because $\mathcal{P}^{(p)}$ is affine in $k$, the parametric constraint can be inverted explicitly. For indemnity, monotonicity of $\mathcal{P}^{(d)}(d)$ in $d$ allows us to invert the constraint numerically (or in closed form under the censored exponential moments used here).

Figure~\ref{fig:budget_combined}(a) fixes $\gamma_d=1,\!000$ (and the remaining parameters) and plots, as functions of $\bar{P}$: (i) the \textit{effective premium} actually spent by each product (which can be strictly below $\bar{P}$ once the unconstrained optimum becomes affordable), and (ii) the corresponding budget-constrained mean--variance values $MV^{(p)}_{\mathrm{bud}}(\bar{P})$ and $MV^{(d)}_{\mathrm{bud}}(\bar{P},\gamma_d)$. For tiny budgets, indemnity may be infeasible due to the additive friction: the minimum feasible indemnity premium at deductible $L$ is $
P_d^{\min}(\gamma_d)
=
\mathcal{P}^{(d)}(L;\theta_d,\gamma_d)
=
(1+\theta_d)\gamma_d .
$
In that region, the best indemnity-or-no-insurance option is simply no insurance ($MV^{(0)}$), while parametric can already begin providing welfare gains if $\gamma_p$ is smaller and the linear premium constraint allows $k>0$. Although parametric payouts may be small relative to extreme losses in this region, their welfare value arises precisely from providing immediate and certain resources when indemnity coverage is infeasible. The comparison is therefore second-best by construction, and the relevant benchmark is the no-insurance outcome rather than full loss replacement.

As $\bar{P}$ increases, parametric mean--variance typically \textit{rises and then flattens} once the unconstrained $k^*$ becomes affordable. Indemnity MV is flat at $MV^{(0)}$ until $\bar{P}$ reaches $P_d^{\min}(\gamma_d)$, and then increases as affordable deductibles begin to reduce variance efficiently. The two MV curves may cross at an \textit{indifference budget} $\bar{P}_{\mathrm{indif}}$: parametric is better for small budgets, but indemnity dominates once a sufficient premium is available. 

To generalize Figure~\ref{fig:budget_combined}(a) across additive indemnity frictions, Figure~\ref{fig:budget_combined}(b) plots the \textit{incremental welfare} conferred by parametric insurance relative to the best alternative among ``indemnity within budget'' and ``no insurance'':
\begin{equation}\label{eq:surf-def}
  \Delta MV(\bar{P},\gamma_d)
  =
  MV^{(p)}_{\mathrm{bud}}(\bar{P})
  -
  MV^{(d)}_{\mathrm{bud}}(\bar{P},\gamma_d),
  \qquad
  \text{shown only when } \Delta MV(\bar{P},\gamma_d)>0.
\end{equation}
The surface therefore appears exactly in the region where parametric insurance is \textit{strictly welfare-improving} relative to the best feasible indemnity choice (including the option of not insuring at all). The surface is absent when $\Delta MV\le 0$, which can occur for two distinct reasons: 
(i) the budget is so small that both products optimally reduce to no insurance (so $\Delta MV=0$), or
(ii) the budget is large enough (and/or $\gamma_d$ small enough) that a deductible contract becomes feasible and dominates parametric
(so $\Delta MV<0$).
Importantly, when indemnity is infeasible because $\bar{P}<P_d^{\min}(\gamma_d)$, we have $MV^{(d)}_{\mathrm{bud}}(\bar{P},\gamma_d)=MV^{(0)}$ by construction, so \textit{parametric can indeed dominate in that region} whenever it can deliver any strict improvement over $MV^{(0)}$.

For any fixed $\gamma_d$ slice of the surface in panel~(b), \eqref{eq:surf-def} reproduces the intuition in panel~(a): the advantage of parametric typically starts at $0$ when $\bar{P}=0$, increases as small budgets allow meaningful parametric payments. At the same time, indemnity remains infeasible or dominated by no insurance, and then declines once indemnity becomes feasible and (eventually) overtakes parametric at higher budgets.

\section{Policy implications}

Our results have implications for disaster risk financing, insurance regulation, and the design of public--private risk-sharing mechanisms in high-risk environments. While the analysis is stylized, it highlights structural limitations of indemnity insurance in environments where affordability and operational frictions bind --- precisely the settings in which risk transfer is most socially valuable. The results should be read as carving out the conditions under which \textit{partial} risk transfer via parametric instruments becomes economically meaningful, rather than as a general claim against indemnity cover.

\subsection{Insurance failure in high-risk areas}

A growing body of evidence indicates that indemnity insurance is becoming unaffordable or unavailable in regions exposed to severe natural hazards, such as floods, wildfires, and cyclones. In these areas, insurers either raise premiums sharply, impose large deductibles, or withdraw coverage altogether, as recently observed in parts of California’s wildfire market and in flood-prone regions worldwide.

Our analysis provides a microeconomic foundation for this phenomenon. Even when indemnity insurance remains theoretically optimal under expected utility maximization, realistic pricing frictions---such as fixed costs, higher capital requirements, and binding budget constraints---push optimal deductibles to levels that render coverage ineffective in practice. In such cases, indemnity insurance ceases to perform its primary economic function: smoothing consumption in adverse states of the world.

This distinction between \textit{theoretical optimality} and \textit{economic relevance} is crucial. A deductible that is optimal in a frictionless model may be so large that it offers little or no meaningful protection to households facing liquidity constraints. From a policy perspective, such outcomes should be interpreted as market failure, even if insurers remain solvent and contracts remain actuarially fair.

\subsection{Parametric insurance as partial risk transfer}

Within this context, parametric insurance emerges as a form of \textit{partial risk transfer} that can dominate indemnity insurance once realistic constraints are acknowledged. While parametric contracts introduce basis risk and do not replicate loss-contingent indemnification, they avoid several frictions inherent to indemnity insurance.

First, parametric payouts can be structured to remain affordable even when indemnity premiums escalate. Because the insurer does not bear loss severity risk, capital requirements and associated loadings are lower \citep{Cummins2012}. Second, parametric contracts remain economically meaningful at low premium levels: even modest budgets translate into immediate protection, whereas indemnity insurance may offer no effective coverage until losses exceed a very high deductible.

Our results show that, under budget constraints, parametric insurance can deliver higher expected utility than the best feasible indemnity contract, even for risk-averse agents. This finding does not overturn the classical optimality of indemnity insurance in frictionless settings \citep{Arr63,Raviv1979}; rather, it clarifies the economic conditions under which that result ceases to be policy-relevant.

\subsection{Speed of payout, recovery, and resilience}

An important policy dimension---only imperfectly captured in our utility framework---is the timing of payouts. Indemnity insurance typically involves claims assessment, verification, and reconstruction processes that can take months or years, particularly following large-scale disasters when local capacity is constrained \citep{Kahn2021}.

Parametric insurance, by contrast, allows for rapid disbursement of funds, often within days or weeks of the triggering event \citep{ClarkeDercon2016}. Even if payouts are smaller or imperfectly aligned with actual losses, early liquidity can materially improve recovery outcomes. This suggests that the welfare gains from parametric insurance identified in our model may understate its broader social value.

Furthermore, parametric insurance at the household level could also potentially mirror the logic of fiscal  risk layering in sovereign disaster-risk management. Governments routinely rely on parametric or 
quasi-parametric instruments not to cover the full cost of disasters, but to secure rapid liquidity 
that stabilizes short-run fiscal and economic conditions. Rather than a replacement, parametric insurance could function as a liquidity layer that complements, rather than replaces, other available risk transfers.

From a public finance perspective, faster payouts may also reduce reliance on ad hoc government assistance and emergency relief, thereby lowering fiscal uncertainty and political pressure following disasters.

\subsection{Reconstruction incentives and resilience investment}
\label{subsec:bbb}

Indemnity insurance typically reimburses losses on a ``like-for-like'' basis, replacing damaged assets with their pre-disaster equivalents. While this aligns with traditional notions of indemnification, it has long been recognized that this approach can fail to reward investment in more resilient reconstruction: improvements that reduce future vulnerability (such as flood-resilient materials, elevated structures, or non-return valves) may not be fully covered if they exceed the cost of restoring the original asset.

In response, indemnity-based systems have developed dedicated ``Build Back Better'' (BBB) mechanisms, following Priority~4 of the Sendai Framework for Disaster Risk Reduction \citep{UNDRR2015}. The UK Flood~Re BBB scheme \citep{FloodRe2022}, launched in 2022, reimburses participating insurers up to \pounds10{,}000 in property flood-resilience measures \textit{in addition to} the cost of like-for-like repair. In Australia, Suncorp Insurance's ``Build it Back Better'' program \citep{Suncorp2025} provides up to AUD\,15{,}000 in resilience enhancements (cyclone shutters, gutter guards, raised utilities, etc.) for substantially damaged homes. More broadly, recent EIOPA/ECB work documents a diversity of European public--private natural-catastrophe schemes in which risk mitigation incentives are integrated into the indemnity architecture \citep{EIOPAECB2023, EIOPAECB2024}. A common feature of all these mechanisms is that they rely on the indemnity claims process: the loss-adjuster visit is the enabling moment at which resilience measures are specified, surveyed, and reimbursed.

Parametric insurance does not, in itself, replicate these mechanisms. Because payouts are not tied to a loss-adjuster's inspection, there is no natural institutional moment at which to specify, certify, or reimburse resilience measures. Households retain discretion over how funds are used. This provides \textit{flexibility}, but it is difficult to deliver systematic resilience investments without further policy scaffolding.

While parametric insurance \textit{removes} the (usual) indemnity constraint that payouts be tied to original-like repair, it does not, on its own, \textit{create} the positive incentive to rebuild better. Hence, realizing resilience gains in a parametric framework would require complementary institutional design. These could include matching public resilience funds, conditional grants, rebuilding guidelines, or tax incentives for adaptation \citep{Hallegatte2019}. Such design, however, is beyond the scope of this paper. Parametric and indemnity insurance are, in this respect, complementary rather than substitutes: hybrid (or layered) architectures combining a parametric liquidity layer with a smaller indemnity component, complemented by an indemnity-based cover that retains the loss-adjuster interface, are a natural avenue for future work. At the sovereign and portfolio
level, an analogous question of how to optimally layer parametric insurance alongside reserve funds and contingent credit has been recently formalized
by \citet{CaSa26}.

\subsection{Demand-side barriers to adoption}\label{subsec:barriers}

The welfare comparison above assumes that households can evaluate a parametric contract on its risk--return merits. In practice, several demand-side frictions lie outside our utility framework but materially affect whether the welfare gains we identify can be realized.

First, \textit{consumer understanding and perceived fairness}. Industry surveys and supervisory reviews consistently identify limited familiarity with parametric products as a leading barrier to household adoption \citep{BISFSI2024, ReinsuranceNews2024}: policyholders must reason about the probability that an index will trigger (somewhat) \textit{independently} of the probability that they will suffer a loss, a cognitively demanding distinction. The behavioral counterpart of basis risk is also distinct from its welfare cost: ex-post outcomes in which a policyholder suffers a loss but receives no payout (or vice versa) have been shown to erode trust in index insurance and depress renewal rates \citep[e.g.,][]{Clarke2016, CaJaSaSa17}, even when such outcomes are ex-ante utility-maximizing. Transparent trigger definitions and explicit disclosure of basis risk are therefore preconditions for take-up.

That being said, continuing on the motivating example of flood insurance considered in this paper, people living in a river basin subject to regular floods would be very familiar with the potential consequences of the river rising to various levels, and could arguably be quite comfortable with a trigger defined on that basis, especially if the cover is only one of several (such as in hybrid or layered approaches discussed above). Furthermore, parametric covers such as provided by FloodFlash trigger based on a gauge attached to the insured's house, which is easy to understand \citep[see also][who discuss the potential benefits of parametric insurance in the flood prone Lismore area of New South Wales, Australia]{JaMeMa24}.

Second, \textit{scale and diversification economies in pricing}. Loading factors $\theta_p$ decline with the size and geographic diversity of the insured population: in a small nascent market, idiosyncratic risk may not diversify away as readily, and required capital per policy is higher. For parametric catastrophe cover this effect is particularly pronounced because triggers tend to be \textit{systemic} --- a single cyclone simultaneously triggers every policy in its footprint --- so pooling across policyholders within a region may deliver less diversification than indemnity cover. This is because while indemnity covers suffer from the same concentration risk with respect to frequency, severity remains (somewhat) idiosyncratic (but weakly so, as a severe event would affect all insured in the area in a similar, negative way). Nascent household parametric markets may therefore face temporarily elevated $\theta_p$, shrinking the region in which our welfare advantage applies.

Third, \textit{regulatory and accounting frictions}. In several jurisdictions, parametric contracts sit ambiguously between insurance and derivative contracts for regulatory and accounting purposes, creating uncertainty for insurers, brokers, and policyholders \citep{BISFSI2024}. Traditional distribution channels are more familiar with indemnity products, and consumer-protection rules designed for indemnity cover may not map cleanly onto parametric designs.

Taken together, these frictions may narrow the welfare region in which parametric insurance dominates.

\subsection{Implications for insurance regulation}

Our findings suggest that insurance regulation focused exclusively on indemnity products may be ill-suited for high-risk environments. Regulatory frameworks that implicitly privilege indemnity insurance---through capital rules, consumer protection standards, or product approval processes---may inadvertently suppress welfare-enhancing alternatives.

A more flexible regulatory approach would recognize parametric insurance as a distinct class of risk-transfer instruments rather than a second-best substitute for indemnity. This includes clarifying disclosure requirements around basis risk, ensuring contract transparency, and facilitating experimentation with hybrid designs that combine parametric triggers and partial indemnification \citep{HuSh24}.

Importantly, the results caution against policies that seek to preserve the affordability of indemnity insurance at all costs, for example, through premium caps or cross-subsidies that obscure underlying risk. In settings where full indemnification is no longer economically viable, encouraging partial but
effective risk transfer may be preferable to maintaining the appearance of comprehensive coverage.

\subsection{The role of government in enabling parametric markets}

Finally, our analysis underscores a constructive role for governments in supporting parametric insurance markets without directly providing insurance. Governments can contribute by investing in high-quality hazard data, standardized indices, and transparent trigger definitions, all of which reduce
basis risk and transaction costs \citep{WorldBank2021}.

Recent Australian reviews illustrate this direction of travel: proposals have emphasized parametric options for household catastrophe protection alongside investments in hazard measurement and transparent triggers \citep{JaMeMa24}.

Rather than crowding out private insurance, well-designed parametric schemes can complement public disaster risk management by strengthening household
resilience and reducing long-term fiscal exposure. In this sense, parametric insurance should be viewed not as a replacement for indemnity insurance, but
as a policy-relevant addition to the menu of risk-transfer instruments in a changing climate.

\section{Conclusion}

This paper revisits the classical optimality of indemnity insurance in environments characterized by low-frequency, high-severity risk and binding budget constraints. While excess-of-loss indemnity contracts are well known to maximize expected utility under proportional premium loadings, we show that this benchmark can lose practical relevance when fixed costs, heterogeneous loadings, and affordability constraints are taken into account. In such settings, the deductible implied by utility maximization may become so large that indemnity insurance delivers little effective risk transfer, despite remaining formally optimal.

Within a tractable expected mean--variance framework, we compare indemnity and parametric insurance designs under a common objective and shared trigger. Allowing for realistic frictions that penalize indemnity insurance---including higher loadings, fixed costs, and delayed loss adjustment---we identify regions in which parametric contracts yield strictly higher welfare. These gains arise not because parametric insurance dominates indemnity insurance in general, but because it remains feasible and economically meaningful precisely when indemnity insurance becomes unaffordable or ineffective.

Our results highlight a non-monotonic welfare comparison between the two designs. Parametric insurance may dominate at low premium budgets, lose its advantage as indemnity insurance becomes viable, and eventually become irrelevant once both contracts are unconstrained. This pattern helps reconcile classical insurance theory with the observed resurgence of parametric risk transfer in high-risk environments.

The analysis has direct implications for disaster risk financing and insurance market design. As climate change and urban development expand the set of regions where full indemnification is no longer viable, policies that focus exclusively on restoring traditional indemnity insurance may be insufficient. Parametric insurance should instead be viewed as a complementary instrument that can improve individual welfare and resilience when standard contracts fail. Future research could extend the framework to richer utility specifications, endogenous risk mitigation, and hybrid insurance designs that combine parametric and indemnity elements. Because neither design dominates globally, such hybrid contracts --- combining a parametric liquidity layer with a reduced indemnity component --- are a natural avenue for future work and for product innovation, particularly for those policyholders facing binding affordability constraints; the analogous \textit{sovereign-level} problem of optimal layering across reserve funds, contingent credit, and parametric
insurance is treated in concurrent work by \citet{CaSa26}.

\section*{Acknowledgements}

This paper was presented at the First ASTIN Bulletin Conference in Zurich in January 2026. The authors are grateful to attendees, as well as to anonymous reviewers, whose comments led to significant improvements of the paper.

Avanzi acknowledges support under the Australian Research Council's Discovery Project (DP200101859) funding scheme. The views and opinions expressed in this paper are solely those of the authors and do not reflect those of their affiliated institutions.

\section*{Data availability statement}

All numerical results and figures in this paper are reproducible using R, with code available on GitHub at \url{https://github.com/agi-lab/parametric}.

\section*{Use of AI tools}
The authors used large language models (ChatGPT, OpenAI; Claude, Anthropic) to assist with drafting and refining portions of the exposition and mathematical derivations. All results, interpretations, and final wording are the authors’ own, and the authors take full responsibility for the content of the paper.

\section*{References}

\bibliographystyle{elsarticle-harv}
\bibliography{libraries2}

\newpage

\appendix

\section{Moments and mean--variance objectives under censored exponential severity}
\label{App:MV_censored_exp}

This appendix collects the moment formulas and the mean--variance objectives
used throughout the paper, under the assumption that the number of events
$N \sim \mathrm{Pois}(\lambda)$ and that severities are i.i.d.\ and censored at
$L>0$. Specifically, if $Z\sim\mathrm{Exp}(\nu)$ then our losses $Y_i$ are distributed as a censored version of $Z$, that is, 
\[
Y_i \sim \min\{Z,L\}, \quad i=1, \ldots.
\]
Then $Y_i$ has density $\nu e^{-\nu y}$ on $[0,L)$ and an atom of mass
$e^{-\nu L}$ at $L$, i.e.
\[
dF_{Y_i}(y) = \nu e^{-\nu y}\,dy + e^{-\nu L}\,\delta_L(dy), \qquad 0 \le y \le L.
\]
We denote by $S=\sum_{i=1}^N Y_i$ the aggregate annual loss.

\subsection{Basic severity moments}
\label{App:MV_censored_exp:basicmom}

The first two raw moments of $Y_i$ are
\begin{align}
\mathbb{E}[Y_i]
&= \int_0^L y \nu e^{-\nu y}\,dy + L e^{-\nu L}
 = \frac{1 - e^{-\nu L}}{\nu},
\label{eq:EY_cens}
\\[4pt]
\mathbb{E}[Y_i^2]
&= \int_0^L y^2 \nu e^{-\nu y}\,dy + L^2 e^{-\nu L}
 = \frac{2}{\nu^2} - \frac{2 e^{-\nu L}}{\nu^2} - \frac{2L e^{-\nu L}}{\nu}.
\label{eq:EY2_cens}
\end{align}
Hence
\begin{equation}
\mathrm{Var}(Y_i)=\mathbb{E}[Y_i^2]-\bigl(\mathbb{E}[Y_i]\bigr)^2.
\label{eq:VarY_cens}
\end{equation}

\subsection{Excess--loss moments for the deductible contract}
\label{App:MV_censored_exp:excessmom}

For a deductible level $d\in[0,L]$, define $(x)_+:=\max\{x,0\}$. Then
\begin{align}
\mathbb{E}\bigl[(Y_i-d)_+\bigr]
&= \int_d^L (y-d)\nu e^{-\nu y}\,dy + (L-d)e^{-\nu L}
 = \frac{e^{-\nu d}-e^{-\nu L}}{\nu},
\label{eq:EYminusd_cens}
\\[4pt]
\mathbb{E}\bigl[(Y_i-d)_+^2\bigr]
&= \int_d^L (y-d)^2\nu e^{-\nu y}\,dy + (L-d)^2e^{-\nu L}
 = \frac{2 e^{-\nu d}}{\nu^2} - \frac{2 e^{-\nu L}}{\nu^2}
 + \frac{2(d-L)e^{-\nu L}}{\nu}.
\label{eq:EYminusd2_cens}
\end{align}
We also record the mixed moment
\begin{align}
\mathbb{E}\bigl[Y_i (Y_i-d)_+\bigr]
&= \int_d^L y(y-d)\nu e^{-\nu y}\,dy + L(L-d)e^{-\nu L}
\notag\\
&= \frac{d(e^{-\nu d}+e^{-\nu L})}{\nu}
   + \frac{2 e^{-\nu d}}{\nu^2}
   - \frac{2 e^{-\nu L}}{\nu^2}
   - \frac{2L e^{-\nu L}}{\nu}.
\label{eq:EY_Yminusd_cens}
\end{align}

A simplification that is repeatedly used is
\begin{align}
G(d):&=\mathbb{E}[\min(Y_i,d)^2]=\mathbb{E}[(Y_i-(Y_i-d)_+)^2]
\notag\\
&= \mathbb{E}[Y_i^2] + \mathbb{E}\bigl[(Y_i-d)_+^2\bigr]
-2\mathbb{E}\bigl[Y_i(Y_i-d)_+\bigr]
\notag\\
&= \frac{2}{\nu^2}\bigl[1-e^{-\nu d}(1+\nu d)\bigr],
\qquad d\in[0,L].
\label{eq:simplification_Gd}
\end{align}

\subsection{Premiums under the expectation principle}
\label{App:MV_censored_exp:premiums}

Under the expectation principle with loading and a fixed-cost term, the
premiums are:
\begin{align}
\prem{d}{d}{\theta_d,\gamma_d}
&= (1+\theta_d)\Bigl(\mathbb{E}\bigl[\Ben{d}{N,Y_1,\dots,Y_N,d}\bigr]+\gamma_d\Bigr)
= (1+\theta_d)\Bigl(\lambda\,\mathbb{E}\bigl[(Y_i-d)_+\bigr]+\gamma_d\Bigr)
\notag\\
&= (1+\theta_d)\left(\lambda\frac{e^{-\nu d}-e^{-\nu L}}{\nu}+\gamma_d\right),
\qquad d\in[0,L],
\label{eq:prem_d_app}
\\[6pt]
\prem{p}{k}{\theta_p,\gamma_p}
&= (1+\theta_p)\Bigl(\mathbb{E}\bigl[\Ben{p}{N,k}\bigr]+\gamma_p\Bigr)
= (1+\theta_p)(\lambda k+\gamma_p),
\qquad k\ge 0.
\label{eq:prem_p_app}
\end{align}

\subsection{Mean--variance objective: deductible indemnity}
\label{App:MV_censored_exp:MVd}

Let terminal wealth be
\begin{equation}
W = w_0 - \prem{.}{.}{.} - S + \Ben{.}{N,Y_1,\dots,Y_N,\cdot},
\end{equation}
and associated mean-variance
\begin{equation}
\MV{.}{W}{\beta} = \mathbb{E}[W]-\beta\,\mathrm{Var}(W),\quad \beta>0.
\end{equation}
For the deductible contract,
\[
\Ben{d}{N,Y_1,\dots,Y_N,d}=\sum_{i=1}^N (Y_i-d)_+,
\qquad
S=\sum_{i=1}^N Y_i.
\]
Using $N\sim\mathrm{Pois}(\lambda)$ and independence of $N$ and $(Y_i)$, one obtains
\begin{equation}
\MV{d}{W}{d,\beta,\theta_d,\gamma_d}
= w_0 - \prem{d}{d}{\theta_d,\gamma_d}
  - \lambda\,\mathbb{E}[Y_i]
  + \lambda\,\mathbb{E}\bigl[(Y_i-d)_+\bigr]
  - \beta\,\lambda\,G(d).
\label{eq:MVd_app}
\end{equation}

Differentiating \eqref{eq:MVd_app} yields
\begin{align}
\frac{\partial \text{MV}^{(d)}}{\partial d}
&= \lambda e^{-\nu d}\bigl(\theta_d-2\beta d\bigr),
\label{eq:MVd_der1}
\\
\frac{\partial^2 \text{MV}^{(d)}}{\partial d^2}
&= \lambda e^{-\nu d}\bigl(-\nu\theta_d+2\nu\beta d-2\beta\bigr).
\label{eq:MVd_der2}
\end{align}
Thus, the interior critical point is
\begin{equation}
d^*=\frac{\theta_d}{2\beta},
\label{eq:dstar_app}
\end{equation}
and the curvature at $d^*$ is negative:
\[
\frac{\partial^2 \text{MV}^{(d)}}{\partial d^2}\Big|_{d=d^*}
= -2\beta\lambda e^{-\nu d^*}<0.
\]
(When contractual bounds are imposed, one projects $d^*$ onto $[0,L]$.)

\subsection{Mean--variance objective: per--event parametric cover}
\label{App:MV_censored_exp:MVp}

For the parametric contract,
\[
\Ben{p}{N,k}=kN,
\qquad
S=\sum_{i=1}^N Y_i.
\]
Under the same assumptions,
\begin{align}
\MV{p}{W}{k,\beta,\theta_p,\gamma_p}
&= w_0 - \prem{p}{k}{\theta_p,\gamma_p}
  - \lambda\,\mathbb{E}[Y_i]
  + \lambda k
  - \beta\lambda\bigl(\mathbb{E}[Y_i^2]+k^2-2k\,\mathbb{E}[Y_i]\bigr).
\label{eq:MVp_app}
\end{align}
Differentiating gives
\begin{equation}
\frac{\partial \text{MV}^{(p)}}{\partial k}
= \lambda\bigl(-\theta_p+2\beta(\mathbb{E}[Y_i]-k)\bigr),
\qquad
\frac{\partial^2 \text{MV}^{(p)}}{\partial k^2}
= -2\beta\lambda<0,
\label{eq:MVp_derivatives}
\end{equation}
so that the unique maximiser is
\begin{equation}
k^*=\mathbb{E}[Y_i]-\frac{\theta_p}{2\beta},
\label{eq:kstar_app}
\end{equation}
(with projection onto any admissible range such as $[0,L]$ if imposed).

\subsection{Optional limiting case as $L\to\infty$}
\label{App:MV_censored_exp:limit}

If $L\to\infty$, then $Y_i\to Y\sim\mathrm{Exp}(\nu)$ with
$\mathbb{E}[Y]=1/\nu$ and $\mathrm{Var}(Y)=1/\nu^2$. In that case,
\eqref{eq:MVp_app} reduces to
\[
\MV{p}{W}{k,\beta,\theta_p,\gamma_p}
= w_0 - (1+\theta_p)(\lambda k+\gamma_p)
  - \lambda\frac{1}{\nu} + \lambda k
  - \beta\lambda\left(\frac{2}{\nu^2}+k^2-\frac{2k}{\nu}\right),
\]
and \eqref{eq:kstar_app} becomes $k^*=\frac{1}{\nu}-\frac{\theta_p}{2\beta}$.
Analogous simplifications apply to the deductible expressions.

\section{Optimal parameters $d^*$ and $k^*$, and conditions for $E[Y_i] = d^* + k^*$}
\label{App:Duality}

We now relax distributional assumptions on $N$ and $Y_i$. This appendix provides full derivations of the optimal deductible $d^{*}$ and the optimal parametric per–event cover $k^{*}$ under the mean–variance objective.  It further identifies the exact conditions under which the duality
\begin{align*}
    E[Y_i] = d^{*} + k^{*}
\end{align*}
holds, and explains why this identity is specific to a combination of 
(i) mean–variance preferences, 
(ii) a Poisson claim count model, and 
(iii) premiums calculated via the expectation principle.  
If any one of these three ingredients is altered, the duality fails.

Let $N$ be the claim count, $Y_i$ the capped severity, 
and $m(d) = E[(Y_i - d)_{+}]$ the expected excess loss above $d$.
Premiums under the expectation principle are
\begin{align*}
    \prem{d}{d^*}{\theta_d,\gamma_d}  = (1+\theta_d)\big( E[N]\, m(d) + \gamma_d \big),
\qquad
\prem{p}{k}{\theta_p,\gamma_p} = (1+\theta_p)\big( E[N]\, k + \gamma_p \big).
\end{align*}

\subsection{General claim count model}

Define $\mu = E[N]$ and $\sigma_N^2 = \Var(N)$, without assuming that $N$ is 
Poisson.  Standard random–sum identities yield, for the parametric cover,
\begin{align*}
    \Var(S - kN)
   = \mu \Var(Y_i) 
     + \sigma_N^2 (E[Y_i] - k)^2.
\end{align*}
The mean–variance objective, therefore, takes the form
\begin{align*}
\MV{p}{W}{k,\beta,\theta_p,\gamma_p}
 = \text{const}
 - \mu \theta_p k
 - \beta\!\left[
      \mu\Var(Y_i)
    + \sigma_N^2(E[Y_i] - k)^2
   \right],
\end{align*}
where 'const' represents a constant in $k$. Differentiation gives
\begin{equation}
k^{*}
 = E[Y_i] 
   - \frac{\mu}{\sigma_N^2}\,\frac{\theta_p}{2\beta}.
 \label{eq:k-general}
\end{equation}
Thus the optimal $k^{*}$ depends on the ratio $\mu/\sigma_N^2$. In particular, 
$k^{*}$ reduces to a simple expression only when $\Var(N)=E[N]$.

For the deductible cover, the retained loss per event is $R_d = \min(Y_i,d)$.
Using random–sum variance formulas,
\begin{align*}
\Var\!\Big(\sum_{i=1}^N R_d\Big)
  = \mu\,\Var(R_d) + \sigma_N^2 (E[R_d])^{2},
\end{align*}
and differentiating the resulting mean–variance objective gives the first–order 
condition
\begin{equation}
\mu\theta_d
  = 2\beta\big(\,\mu d^{*} + (\sigma_N^2-\mu) E[R_{d^{*}}]\,\big),
  \label{eq:d-general}
\end{equation}
which in general has no closed form and depends explicitly on $\Var(N)$.

\subsection{Specialization to the compound Poisson case}

Now assume $N \sim \mathrm{Pois}(\lambda)$, so that 
\begin{align*}
\mu = \sigma_N^2 = \lambda.
\end{align*}
This equi-dispersion property causes significant simplifications. The parametric solution reduces to
\begin{align*}
k^{*}
 = E[Y_i] - \frac{\theta_p}{2\beta}.
\end{align*}

Similarly, for the deductible, the variance identity reduces to
\begin{equation}
\Var\!\Big(\sum_{i=1}^N R_d\Big)
   = \lambda E[R_d^2],
\end{equation}
and the nonlinear term in the first-order condition disappears, such that
\begin{align*}
\lambda\theta_d = 2\beta\lambda d^{*},
\qquad\text{so that}\qquad
d^{*} = \frac{\theta_d}{2\beta}.
\end{align*}

These closed–form solutions rely critically on:
\begin{enumerate}
    \item the mean–variance objective (quadratic variance term),
    \item Poisson equi–dispersion ($\Var(N)=E[N]$), and
    \item expectation–principle premiums (linearity in $E[( Y_i-d)_+]$ 
          and in $k$).
\end{enumerate}
Removing any of these assumptions breaks the linear structure of the 
first–order conditions.

\subsection{Duality}

Assuming identical loadings ($\theta_d=\theta_p=\theta$), we have shown:
\begin{equation}
d^{*} = \frac{\theta}{2\beta},
\qquad
k^{*} = E[Y_i] - \frac{\theta}{2\beta}.
\end{equation}
Hence,
\begin{equation}
E[Y_i] = d^{*} + k^{*}.
  \label{eq:duality_app}
\end{equation}
This identity provides the duality described in Section~3.1.  It is a direct 
consequence of the combination of assumptions (1)--(3) above and does not hold 
for general claim count distributions or for premium principles other than the 
expectation principle.

\end{document}